\documentclass[twocolumn,aps,10pt,pra,nofootinbib,showpacs,superscriptaddress]{revtex4-1}

\usepackage{graphicx}
\usepackage{bm}
\usepackage{amsmath,hyperref,amsthm,ulem}
\usepackage{amssymb}
\usepackage{appendix}
\usepackage{amsfonts}
\usepackage{fancyhdr}
\usepackage{slashed}
\usepackage{latexsym,epsfig,bbm}
\usepackage{qcircuit}

\usepackage{pdfcomment}

\usepackage[inline]{enumitem}

 \theoremstyle{plain}

\newcommand{\bra}[1]{\langle{#1}|}
\newcommand{\ket}[1]{|{#1}\rangle}

\newcommand{\braket}[2]{\langle{#1}|{#2}\rangle}

\usepackage{color}
\definecolor{blue}{rgb}{0,0.2,1}

\definecolor{red}{rgb}{0.9,0,0}

\newcommand{\Ord}[1]{\mathcal{O}\left( #1 \right)}
\newcommand{\tOrd}[1]{\tilde{\mathcal{O}}\left( #1 \right)}

\begin{document}

\title{Photonic quantum algorithm for Monte Carlo integration}
\author{Patrick Rebentrost}
\email{pr@patrickre.com}
\author{Brajesh Gupt}
\email{brajesh@xanadu.ai}
\author{Thomas R. Bromley}
\email{tom@xanadu.ai}
\affiliation{Xanadu, 372 Richmond St W, Toronto, M5V 2L7, Canada}
\date{\today}

\begin{abstract}
We present a continuous-variable photonic quantum algorithm for the Monte Carlo evaluation of multi-dimensional integrals. Our algorithm encodes $n$-dimensional integration into $n+3$ modes and can provide a quadratic speedup in runtime compared to the classical Monte Carlo approach. The speedup is achieved by developing a continuous-variable adaptation of amplitude estimation. We provide an error analysis for each element of the algorithm and account for the effects of finite squeezing. Our findings show that Monte Carlo integration is a natural use case for algorithms using the continuous-variable setting.
\end{abstract}
\maketitle 

\section{Introduction}

Monte Carlo (MC) methods are an important computational approach in many fields of science and technology. One common problem solved through MC methods is the numerical integration of a function. Here, the integrand is evaluated at randomly selected points and the values are averaged to obtain an approximation to the integral.  This procedure  takes a number of evaluations of the integrand
that scales inversely to the square of the desired accuracy. Quantum algorithms have the potential to improve this error scaling, e.g., such that a shorter runtime is required to achieve the same error as classical MC.

A quantum algorithm for database search was first presented by Grover \cite{Grover1996}, 
obtaining a quadratic speedup in the number of queries to an unstructured database for finding a particular element. It was generalized to amplitude amplification in \cite{Brassard2002} and extended to amplitude estimation in the same reference.
Amplitude estimation provides a useful starting point for quantum versions of MC.
In the qubit setting, quantum MC algorithms were discussed, e.g.,~in \cite{Montanaro2015,Xu2018}.
Ref.~\cite{Pati2000} adapts Grover's search algorithm to the quantum continuous-variable (CV) context. To our knowledge, a CV adaptation of amplitude estimation and its extension to Monte Carlo has not so far been considered.

In this work, we introduce a CV version of quantum Monte Carlo (QMC)\footnote{We note that ``quantum Monte Carlo" is commonly used as a descriptor for the study of quantum systems with classical Monte Carlo techniques, for example in quantum chemistry.} for the evaluation of multi-dimensional integrals. 
As an intermediate step, we also adapt the amplitude estimation algorithm to the CV setting. We discuss the steps of our algorithm and highlight the CV versions of familiar qubit-based transformations, including the controlled rotation and reflection operations. Our analysis contains an account of errors, including inaccuracies due to finite squeezing. The resultant algorithm can give quadratic speedups for evaluating integrals, and we discuss under which conditions a speedup is realized.

By focusing on the CV paradigm of quantum computing, our approach confers a number of advantages in comparison to existing results using qubits for speedups in MC~\cite{Montanaro2015,Xu2018,Rebentrost2018finance}. Fundamentally, the mechanics of CV naturally accommodates the language of integration and does not require any discretization of the integration space, as is the case for qubits. The number of modes used for $n$-dimensional CV QMC is $n + 3$, a quantity that is independent of the desired accuracy of integration. However, the ability to squeeze and apply the required cubic phase gates presents a challenge for physical implementations of CV QMC. We summarize the setting of the present work in section \ref{sectionSetting}. We detail each stage of our algorithm for integration over a single dimension in Secs.~\ref{sectionFirstStage} and~\ref{sectionCVMC}, while discussing errors and speedups from the algorithm in Sec.~\ref{sectionErrors}. The extension to multiple dimensions is discussed in Sec.~\ref{secMulti} and an example numerical implementation of phase estimation is given in Sec.~\ref{Sec:Numerics}. We then conclude in Sec.~\ref{sectionDiscussion}.

\section{Setting}
\label{sectionSetting}
Consider a real valued $n$-dimensional function $g(\vec x): \mathbbm R^n \to \mathbbm R$. Its integral over a region $R \subseteq \mathbbm R^n$ is written as
\begin{equation} \label{eqMainInt}
\mathcal{I} := \int_{R} d \vec x \,\, g(\vec x),
\end{equation}
where $\vec x \in \mathbbm R^n$. For many choices of $g(\vec{x})$, the explicit evaluation of $\mathcal{I}$ is hard, and one often resorts to MC sampling to find an approximate solution. 
For many applications, a representation of the integral as an \textit{expectation value} appears naturally, and we focus on this setting in the following. Then 
\begin{equation}\label{eqExpValMain}
\mathcal{I} = \int_{\mathbbm R^n} d\vec x \,\, p(\vec x) f(\vec x),
\end{equation}
where $f(\vec x): \mathbbm R^n \to \mathbbm R$ is another real valued function and $p(\vec x)$ is a multidimensional probability distribution  $p(\vec x): \mathbbm R^n \to \mathbbm R$ with $\int d\vec x \,\, p(\vec x)  = 1$, for example the Gaussian distribution. Here, $f(\vec x)$ is an arbitrary function describing a random variable of the outcomes distributed with $p(\vec x)$, so that $\mathcal{I}$ can be approximated through MC using $N_{C}$ samples as
\begin{equation}
\mathcal{I} \approx \tilde{\mathcal{I}} := \frac{1}{N_{C}}\sum_{\substack{i=1 \\ x_{i} \sim p(x)}}^{N_{C}} f(x_{i}).
\end{equation}
The probability that this approximation is inconsistent beyond an error $\epsilon$ is given by
\begin{equation}\label{eqCheb}
{\rm Pr} \left (|\mathcal{I} - \tilde{\mathcal{I}}| \geq \epsilon \right) \leq \frac{\sigma^{2}}{N_{C} \epsilon^{2}},
\end{equation}
where $\sigma^{2}$ is the variance of $p(x)$. Hence, for a constant error probability it suffices to pick $N_{C} = \mathcal{O}(\sigma^{2} / \epsilon^{2})$.

We introduce a CV quantum algorithm for MC integration and show that it can provide speedups in comparison to the classical approach. Our approach can yield a close to quadratic speedup in processing time with a number of steps $N_Q = \mathcal{O}(1 / \epsilon)$. Our algorithm consists of two stages. The first stage represents computing the integral. Here, an optical mode is prepared following the probability distribution $p(\vec x)$ and a ``controlled rotation" is then enacted with other modes to imprint the random variable $f(\vec x)$. The second stage is to perform a CV version of amplitude estimation that we introduce here, which is achieved by combining with a squeezed resource mode for phase estimation. This stage allows the integral to be extracted with a quadratic speedup in runtime.

In this work, the function $f(\vec x)$ is bounded as $0 \leq f(\vec x) \leq 1$ for all $\vec x$. 
This also means that the desired integral is $\mathcal{I} \leq 1$, since $p(\vec x)$ is a probability density. 
An extension to more general functions was given in the qubit context in Ref.~\cite{Montanaro2015} and the corresponding CV version will be the subject of future work. 

\begin{figure}
$$
\,\,\,\, \Qcircuit @C=1em @R=2em {
& &  \,\,\,\, \,\,\,\, \,\, \mbox{\bf First Stage} & & & & \mbox{\bf Second Stage} \\
& \lstick{\ket{\rm vac}} & \gate{\mathcal{G}} & \multigate{2}{\mathcal{H}} & \qw & \multigate{3}{\mathcal{Q}_{c}} & \multigate{3}{\mathcal{Q}_{c}} & \qw & \ldots & & \multigate{3}{\mathcal{Q}_{c}} & \qw \\
& \lstick{\ket{\rm vac}} & \gate{S} & \ghost{\mathcal{H}} & \qw & \ghost{\mathcal{Q}_{c}} & \ghost{\mathcal{Q}_{c}} & \qw & \ldots & & \ghost{\mathcal{Q}_{c}} & \qw \\
& \lstick{\ket{\rm vac}} & \gate{S} & \ghost{\mathcal{H}} & \qw & \ghost{\mathcal{Q}_{c}} & \ghost{\mathcal{Q}_{c}} & \qw & \ldots & & \ghost{\mathcal{Q}_{c}} & \qw \\
& \lstick{\ket{\rm vac}} & \qw & \qw & \gate{S} & \ghost{\mathcal{Q}_{c}} & \ghost{\mathcal{Q}_{c}} & \qw & \ldots & & \ghost{\mathcal{Q}_{c}} & \meterB{\ket{x}} \gategroup{2}{2}{4}{4}{0.7em}{.} \gategroup{2}{5}{5}{11}{0.7em}{--}
}
$$
	\caption{Quantum circuit diagram for one dimensional integration using CV QMC. Here, $\mathcal{I}=\int dx \,\, p(x) f(x)$ is approximated in two stages. The first stage (Sec.~\ref{sectionFirstStage}) begins by preparing the first mode using $\mathcal{G}$ so that its position wavefunction matches $\sqrt{p(x)}$. The second and third modes are squeezed in the position eigenbasis as much as possible and then $\mathcal{H}$ imprints $f(x)$ using the CV analog of a controlled rotation. Projecting onto regions in the position eigenbasis of the second and third mode then gives $\mathcal{I}$ through the success probability. The second stage (Sec.~\ref{sectionCVMC}) is CV amplitude estimation. Here, multiple applications of the three-mode unitary $\mathcal{Q}$ amplifies the amplitude corresponding to the projection. Adding a squeezed ancilla mode and instead applying the controlled unitary $\mathcal{Q}_{c}$ imprints $\mathcal{I}$ into the final mode. Measuring the final mode in the position eigenbasis then gives the integral as an expectation value. The total number of applications of $\mathcal{Q}_{c}$ can be $\mathcal{O}(1/\epsilon)$ for an approximation error~$\epsilon$. The errors in CV QMC are accounted for in Sec.~\ref{sectionErrors}.}
\label{Fig:Diagram}
\end{figure}
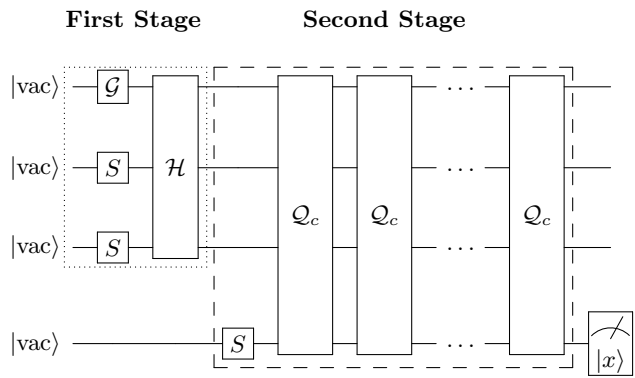

We now proceed to explain CV QMC and then discuss the speedups and errors. The following focuses on the case of one dimensional integration, and we extend to multiple dimensions in Sec.~\ref{secMulti}. Figure~\ref{Fig:Diagram} shows the quantum circuit diagram for one-dimensional integration using CV QMC, requiring $4$ modes and split up into the two stages.

\section{First Stage: Encoding the integral}
\label{sectionFirstStage}
The objective of the first stage of CV QMC is to encode the integral $\mathcal{I}$. This stage uses three modes: the first mode is prepared dependent on $p(x)$ and the other two modes are used to imprint $f(x)$. The probability of successfully postselecting on these two modes then gives the integral $\mathcal{I}$. This method uses a projective measurement so that the CV version of amplitude estimation can be enacted in the second stage of our algorithm.

\subsection{Initial states}
\label{sectionInitial}

In most algorithms of CV quantum computation the initial states are prepared in the vacuum, $\ket{\rm vac}$. As is the case here, it can also be useful to theoretically work with infinitely squeezed $\hat{q}$ (position) eigenstates $\ket {x_{0}}_q $. See Appendix \ref{appendixCVBasics} for an introduction to some elements of continuous-variable quantum computing. Using the squeezing and displacement gates, these states are prepared from the vacuum by infinitely squeezing and then displacing by ${x_{0}}$, i.e.,
\begin{equation}
\ket {x_{0}}_q = \lim_{r\to \infty} D\left (x_{0} \right)S(r) \ket{\rm vac}.
\end{equation}
Realistically, there is a maximum squeezing factor $r_{\max}$ achievable in physical implementations, which introduces errors. For an approximation to $\ket {x_{0}}_q $ we can use a finitely squeezed and displaced coherent state
\begin{eqnarray}
\ket{G_{{x_{0}},s} } &=& D\left (x_{0} \right)S\left (r \right) \ket{\rm vac} \nonumber \\
&=& \int dx \,\, G_{x_{0}, s}(x) \ket{x}_q,
\end{eqnarray}
with a squeezing factor of $r \rightarrow r_{\max}$ and the squeezing $s:=\frac{1}{\sqrt 2} e^{-r}\to s_{\min}$. For these states the wavefunction $G_{{x_{0}}, s}(x)$ is proportional to a Gaussian with a standard deviation proportional to $s$ and a mean ${x_{0}}$. For some of the discussion, we assume availability of position eigenstates $\ket{{x_{0}}}_{q}$ using infinite squeezing and then account for the error effects of using finitely squeezed and displaced coherent states $\ket{G_{{x_{0}},s} }$.

\subsection{Preparing $p(x)$}
\label{sectionG}

The first stage of CV QMC begins with preparing a mode according to the probability distribution $p(x)$. Precisely, we assume availability of a unitary $\mathcal G$ such that
\begin{equation}\label{Eq:Prep}
\mathcal G \ket{\rm vac} = \int dx \sqrt{p(x)} \ket x_q.
\end{equation}
As discussed in Appendix~\ref{appendixDecomposition}, $\mathcal G$ can be implemented by approximating it through a decomposition into a universal set of elementary CV gates, such as Gaussian unitaries and the cubic phase gate. For CV QMC to provide speedups over conventional MC, the decomposition of $\mathcal{G}$ into elementary gates must be efficient, see Sec.~\ref{sectionErrors}.

To highlight an important case, many problems involve the Gaussian probability density
\begin{equation}
p(x) =G_{x_0,\sqrt{2} \sigma}^2(x).
\end{equation}
with $\sigma$ standard deviation and $x_{0}$ mean. The square root of this density is proportional to
$\sqrt{p(x)} \propto G_{x_0,\sqrt 2 \sigma}(x)$. We can then prepare a mode in the state $\ket{G_{x_0, \sqrt 2\sigma}}$ by applying
\begin{equation}
\mathcal{G} = D\left(x_{0} \right) S\left(- \log \sqrt{2} \sigma \right)
\end{equation}
to the vacuum.

\subsection{Applying $f(x)$}
\label{sectionF}

The random variable function $f(x)$ is imprinted by interacting with an additional two ancillary modes. The interaction is given by the three-mode gate
\begin{equation}\label{eqGateHid}
\mathcal H^{\rm id} := e^{-i  \left (1/\sqrt{f(\hat q_1)} \right) \otimes \hat p_2 \otimes \hat p_3}.
\end{equation}
Here, we use the $^{\rm id}$ superscript to denote the ``ideal" version of the gate. 
This gate acts with the function $1/\sqrt{f(\hat q_1)}$ on the first mode, where $\hat q_1$ is the position operator of the first mode, and with the momentum operators $\hat p_2$ and $\hat p_3$ on the ancilla modes. 

To understand the action of $\mathcal{H}^{\rm id}$, we can apply it to $\ket{x}_{q_{1}}\ket{0}_{q_{2}}\ket{0}_{q_{3}}$.  Using $\ket{0}_{q_2} =  \int dp'  \ket{p'}_{p_2} $, the result is
\begin{eqnarray}
 \mathcal H^{\rm id} \ket x_{q_1} \ket{0}_{q_2} \ket 0_{q_3}  
&=& \ket x_{q_1}  \int dp'  \ket{p'}_{p_2} \ket { \frac{ p'}{\sqrt{f(x)}} }_{q_3}.\label{eqApplyH}
\end{eqnarray}
Hence the interaction can be interpreted as the CV analogue of a controlled rotation, 
i.e.,~adding a $\frac{1}{\sqrt{f(x)}}$ position displacement onto the last mode dependent upon the position eigenstate 
$\ket{x}_{q_{1}}$ of the first mode. In the qubit version, an ancilla qubit is rotated by an amount determined 
by another register of qubits, see Appendix \ref{appendixH} for more discussion on this analogy. 

Similarly to the qubit case, we can perform a measurement on the ancillary modes of $\mathcal H^{\rm id} \ket x_{q_1} \ket{0}_{q_2} \ket 0_{q_3}$ to obtain an amplitude encoding of the function $\sqrt{f(x)}$, or the function itself $f(x)$ through the probability of success. We measure the second mode postselected in the state $\ket{0}_{q_2}$ 
and the third mode in the state $\ket{x_{\rm off}}_{q_3}$. The offset value $x_{\rm off}$ is arbitrary and may be chosen according to experimental convenience. 
Applying $\bra{0}_{q_2}\otimes \bra{x_{\rm off}}_{q_3}$ results in
\begin{eqnarray}
&&\left (\bra{0}_{q_2}\otimes \bra{x_{\rm off}}_{q_3}\right ) \mathcal H^{\rm id} \ket x_{q_1} \ket{0}_{q_2} \ket 0_{q_3} 
= \frac{\sqrt{f(x)} }{2\sqrt{\pi}} \ket x_{q_1}, \nonumber\\ \label{eqHPostSelect}
\end{eqnarray}
see Appendix \ref{appendixH} for the intermediate steps.
This means that the resultant state is $\ket{x}_{q_{1}}$ with a probability proportional to $f(x)$. 
 
For a physical implementation of CV QMC, the function $1/\sqrt{f(\hat{q}_{1})}$ can be implemented via a polynomial approximation $h(\hat{q}_{1})$. 
We can in principle implement the exponentiated polynomial $e^{-i  h(\hat q_1)}$ by decomposing it 
into a sequence of Gaussian single-mode gates and cubic phase gates. Such decompositions have been studied previously, for example in \cite{Sefi2011}, and are also discussed in Appendix \ref{appendixDecomposition}. 
This decomposition must be efficient to provide useful speedups through CV QMC, see Sec.~\ref{sectionErrors}.
The three-mode interaction using $h(x)$ is given by
\begin{equation}\label{eqGateHimpl}
\mathcal H \equiv \mathcal H^{\rm impl} := e^{-i  h(\hat q_1)\otimes \hat p_2 \otimes \hat p_3}.
\end{equation}
Here, we use the $^{\rm impl}$ superscript to denote the ``implementation" version of the gate. 
This gate is generated by a higher-order polynomial in the position and momentum operators of the three modes. As the single mode gate $e^{-i  h(\hat q_1)}$, it can also be decomposed into a sequence of single and two-mode Gaussian operations and single-mode cubic phase gates. 
If the decomposition of $e^{-i  h(\hat q_1)}$ is efficient then also the decomposition of $\mathcal H^{\rm impl}$ is efficient. 

This approach is a generalization of the technique used by Lau \textit{et al.} \cite{Lau2016}
for performing a quantum matrix inversion \cite{Harrow2009} in the CV setting. It achieves an encoding of the function $1/\vert h(x)\vert \approx \sqrt{f(x)}$ as an amplitude of the position eigenstate $\ket x_{q_1}$. The interaction can of course be performed on a superposition state in the position eigenbasis, which is the route now used to obtain our desired integral.

\subsection{Obtaining the integral}
\label{sectionIntegral}

We can combine the state preparation of part B with the controlled rotation and postselection on ancilla modes detailed in part C to obtain $\mathcal{I}$. Take the three-mode vacuum state to be the initial state 
\begin{equation}
\ket{\psi_{\rm in}}:=\ket {\rm vac}_{1}  \ket{\rm vac}_{2} \ket{\rm vac}_{3},
\end{equation}
First, define the operator
\begin{equation} \label{eqOpK}
\mathcal K^{\rm id} := \mathcal H^{\rm id} \left (\mathcal G\otimes S(\infty) \otimes S(\infty) \right ).
\end{equation}
This ``ideal" operator consists of preparing the superposition over all position eigenkets in the first mode with amplitudes $\sqrt{p(x)}$  via $\mathcal G$, infinitely squeezing the ancilla modes via $S(\infty)$, and finally applying the three-mode controlled rotation gate $\mathcal{H}^{\rm id}$ that encodes $f(x)$. Finitely squeezed initial states are accounted for in Appendix \ref{App:FiniteSqueezing}. We define the resulting state as
\begin{equation}
\ket {\chi^{\rm id}} := \mathcal K^{\rm id} \ket{\psi _{\rm in}}.
\end{equation}

After preparing the first mode and squeezing the ancillas, we have as an intermediate state 
\begin{equation}
\mathcal G \otimes S(\infty)\otimes S(\infty)\ket{\psi_{\rm in}} = \int dx \sqrt{p(x)} \ket x_{q_1}  \ket{0}_{q_2} \ket 0_{q_3}.
\end{equation}
Then applying $\mathcal H$, see Eq.~(\ref{eqApplyH}), obtains
\begin{equation}
\ket {\chi^{\rm id}} = \int dx \sqrt{p(x)} \ket x_{q_1}  \int dp'  \ket{p'}_{p_2} \ket {\frac{ p'}{\sqrt{f(x)}}}_{q_3}.
\end{equation}
Postselecting on the resource modes in the infinitely squeezed states $\ket{0}_{q_2}\otimes \ket{x_{\rm off}}_{q_3}$, using Eq.~(\ref{eqHPostSelect}), arrives at
\begin{equation}\label{eqHPostSelectSuperposition}
\frac{1}{2\sqrt{\pi} }\int dx \,\, \sqrt{p(x) f(x)}   \ket x_{q_1} .
\end{equation}
The postselection success probability is 
\begin{equation}\label{eqHPostSelectProb}
\frac{1}{4\pi} \int_{-\infty}^{\infty} dx \,\, p(x)f(x) = \frac{\mathcal I}{4\pi},
\end{equation}
which is proportional to the desired integral $\mathcal I$.
 
For the results in Eqs.~(\ref{eqHPostSelect}), (\ref{eqHPostSelectSuperposition}), and (\ref{eqHPostSelectProb}), postselection is performed on infinitely squeezed states. In other words, the operator
\begin{equation}\label{Eq:M}
\mathcal{M} := \mathbbm I\otimes \ket{0}_{q_2} \bra{0}_{q_2} \otimes \ket{x_{\rm off}}_{q_3} \bra{x_{\rm off}}_{q_3}
\end{equation}
is measured, where $\mathbbm I$ is the identity operator. However, it is unphysical to be able to measure such an operator as one can only measure the position in a finite interval with spread  $\Delta x$. The physically realizable $\Delta x$ is larger than the spread required for an ideal operation $\Delta x' \to 0$. We now account for this physical setting by introducing a suitable \textit{projector}.
The following CV amplitude estimation algorithm also requires a projector, while here $ \mathcal{M}^2 \neq  \mathcal{M}$ since the infinitely squeezed states are not normalizable. 
\footnote{In the qubit setting, this requirement can be relaxed, see e.g.~\cite{Xu2018}, and the measurement can be a Hermitian operator. We leave this case for future discussion.}

There are multiple ways to define a projector that obtains the integral to a certain approximation. An alternative projector to the present work is the projector discussed in \cite{Pati2000}, see also Appendix \ref{appendixPBL}. 
In this work, we focus on a projector into squeezed coherent states, defined as
\begin{equation}
\label{eqProj}
P_{x_0,\Delta x} := \ket{G_{x_0,\Delta x} }\bra{G_{x_0,\Delta x} }.
\end{equation}
Note that $P_{x_0,\Delta x}^2 = P_{x_0,\Delta x}$. Such a projector can be measured via the application of (anti-) squeezing and subsequent heterodyne measurement \cite{Weedbrook2012}.
Let the maximum squeezing factor be $r_{\max}$ and the associated squeezing be $s_{\min}$.
In contrast to Eq.~(\ref{Eq:M}), the projector to obtain the desired success probability approximately is given by
\begin{equation}
\mathcal P:=  \mathbbm I\otimes P_{0,s_{\min}} \otimes P_{x_{\rm off},s_{\min}}.
\end{equation}
Note again that $\mathcal P^2 =\mathcal P$. This operator projects into the respective squeezed coherent states with spread $s_{\min}$. 

Before we apply this projector, consider the state $\ket{\chi^{\rm id}}$.
Realistically, we can only apply $\mathcal H^{\rm impl}$ and squeeze with $r_{\max}$, thus applying the operator
\begin{equation} \label{eqOpKimpl}
\mathcal K^{\rm impl} := \mathcal H^{\rm impl} \left (\mathcal G\otimes S(r_{\max}) \otimes S(r_{\max}) \right ),
\end{equation}
which leads to the state 
\begin{equation}
\ket {\chi^{\rm impl}} := \mathcal K^{\rm impl} \ket{\psi _{\rm in}}.
\end{equation}
If we measure this projector on the state $ \ket {\chi^{\rm impl}}$, we obtain
\begin{eqnarray}
\bra {\chi^{\rm impl}} \mathcal P \ket {\chi^{\rm impl}}
&\approx&\frac{s_{\rm min}^4}{ \pi^2 } \mathcal I , \label{eqchiPchi}
\end{eqnarray}
see Appendix \ref{appendixFiniteSqueezing}. 
The measurement returns the integral as before, scaled by the measurement spread
 and the squeezing $s_{\min}$ in the limit of strong squeezing ($s_{\min}\to 0$). 
 The error scales as $\Ord{s_{\min}^6}$, see also Appendix \ref{appendixFiniteSqueezing}. 
The additional error due to the polynomial approximation is discussed in Appendix \ref{appendixErrorPolynomial}. 

To summarize, the integral can be obtained by using a CV analogue of a controlled rotation and then performing a projective measurement (similar to the qubit setting). However, this does not yet provide a speedup in comparison to classical methods since finding $\bra {\chi^{\rm impl}} \mathcal P \ket {\chi^{\rm impl}}$ is achieved experimentally through a simple Bernoulli trial. We now show how a CV version of amplitude estimation can be applied to provide speedups through QMC.

\section{Second stage: Amplitude Estimation and Speedup}
\label{sectionCVMC}

The second stage of CV QMC is to provide a speedup by adding an additional mode and repetitively performing a four mode interaction that encodes the integral as the result of position measurement on the final mode. This stage represents a CV version of amplitude amplification and estimation. 

As discussed in Ref.~\cite{Brassard2002,Knill2007}, amplitude estimation for qubits is a combination of amplitude amplification with quantum phase estimation. We consider both elements in the CV setting. For CV amplitude amplification, we define a continuous-variable operator 
$\mathcal Q$ in analogy to the qubit case (also called the Grover operator in the search context). This operator encodes the desired expectation value in its eigenvalues. CV phase estimation using a single squeezed mode is then performed with a ``controlled" operator $\mathcal Q^c$ to resolve the corresponding eigenvalue.

\subsection{Amplitude amplification}

First, along the lines of \cite{Brassard2002}, we would like to turn the measurement operator into a unitary operator.
Consider the idealized operator 
\begin{equation}\label{eqV}
\mathcal V^{\rm id}  := \mathbbm I^{\otimes 3} - 2 \mathcal M. 
\end{equation}
This operator is not unitary as $\mathcal M$ is not a projector.
Nevertheless, a measurement of $\mathcal V^{\rm id} $ on $\ket {\chi^{\rm id}}$
extracts the desired integral via
$\bra {\chi^{\rm id}} \mathcal V^{\rm id}  \ket {\chi^{\rm id}} =1- \mathcal I/2\pi$.
To obtain a unitary operator we use the previously defined projector
\begin{equation}
\mathcal V^{\rm impl}  := \mathbbm I^{\otimes 3} - 2 \mathcal P. 
\end{equation}
which leads to
$\bra {\chi^{\rm impl}} \mathcal V^{\rm impl}  \ket {\chi^{\rm impl}} \approx 1 - 2 \frac{s_{\min}^4}{ \pi^2 } \mathcal I$.
Recall again the errors due to finite squeezing and polynomial approximations, as discussed further in Sec.~\ref{sectionErrors}.

Using the ideal states for the moment, we can formally express $\mathcal V^{\rm id} \ket {\chi^{\rm id}}$ as 
 a linear combination of $\ket {\chi^{\rm id}}$ and a particular orthogonal complement $\ket {\chi^{{\rm id},\perp}}$, i.e.,
 \begin{equation}
 \mathcal V^{\rm id} \ket {\chi^{\rm id}} = \cos (\theta/2) \ket {\chi^{\rm id}} + e^{i\phi} \sin  (\theta/2) \ket {\chi^{{\rm id},\perp}}.
 \end{equation}
 It follows that
 \begin{equation} \label{eqTheta}
\cos  \left(\frac{\theta}{2}\right)=  1 - \frac{\mathcal I}{2\pi},
 \end{equation}
and we can equivalently think of $\theta$ as containing the desired integral.

Next we use the fact that $\mathbbm{I} - 2 \ket{\psi}\bra{\psi}$ defines a unitary reflection around any state $\ket{\psi}$, i.e., so that $\ket{\psi} \rightarrow - \ket{\psi}$ and $\ket{\psi^{\perp}} \rightarrow \ket{\psi^{\perp}}$ for any orthogonal states. From this, define $\mathcal Q^{\rm id}$ as the ideal operator for amplitude amplification, given by a sequence of a reflection of $\mathcal V^{\rm id} \ket {\chi^{\rm id}}$ followed by a reflection of $\ket {\chi^{\rm id}}$,
 \begin{equation}\label{eqQ}
 \mathcal Q^{\rm id} := \left (\mathbbm I -2 \ket {\chi^{\rm id}} \bra {\chi^{\rm id}}\right )\left(\mathbbm I -2 \mathcal V^{\rm id} \ket {\chi^{\rm id}} \bra {\chi^{\rm id}} \mathcal V^{\rm id}\right ).
 \end{equation}
This operator performs a rotation by an angle of $2\theta$ in the two-dimensional Hilbert space spanned by $\ket{{\chi^{\rm id}}}$ and $\mathcal V^{\rm id}\ket{{\chi^{\rm id}}}$
~\cite{Brassard2002,Knill2007,Rebentrost2018finance}. 
We can hence diagonalize $\mathcal{Q}^{\rm id}$ in this subspace, see Appendix \ref{appendixQ}. This leads to the eigenstates $\ket{ \psi_\pm}$ with corresponding eigenvalues $e^{\pm i \theta}$.

Now, the state $\ket {\chi^{\rm id}}$ outputted from the first stage of CV QMC is the initial state for amplitude estimation. We can express \cite{Xu2018}
\begin{eqnarray}
 \ket {\chi^{\rm id}} = \frac{1}{\sqrt 2} \left ( \ket {\psi_+} +  \ket {\psi_-} \right).
\end{eqnarray}
Applying $\mathcal{Q}^{\rm id}$ to $\ket {\chi^{\rm id}}$ will thus add a phase based on the eigenvalues $e^{\pm i \theta}$. Amplitude amplification consists of repeatedly applying $\mathcal{Q}^{\rm id}$ to $\ket {\chi^{\rm id}}$. The extension to amplitude \textit{estimation}  is to combine with CV phase estimation to imprint both values $\pm \theta$ onto another ancilla mode.  The eigenvalues can then be extracted via homodyne measurement statistics. 

Before proceeding to phase estimation, we first discuss how $\mathcal{Q}^{\rm id}$ can be implemented. Using the definition of $\ket {\chi^{\rm id}} = \mathcal K^{\rm id} \ket{\psi_{\rm in}}$, we can expand the operator into the following sequence of operations
\begin{equation}\label{Eq:Q}
\mathcal Q^{\rm id} = \mathcal K \mathcal Z \mathcal K^\dagger \mathcal V \mathcal K \mathcal Z \mathcal K^\dagger \mathcal V,
\end{equation}
omitting the superscript $^{\rm id}$ for clarity.
The operator $\mathcal Z$ is defined as the reflection of the computational zero state
\begin{equation}\label{eqZ}
\mathcal Z := \mathbbm I^{\otimes 3}  - 2 \ket {\psi_{\rm in }} \bra{\psi_{\rm in}}.
\end{equation}
Implementation of the reflection operators $\mathcal Z$ and $\mathcal V$ is discussed in Appendix~\ref{SecReflec}. This requires invoking a reflection gate of the vacuum state, which is described in Appendix \ref{appendixVacuum}.
The explicit gate sequence of $\mathcal Q^{\rm id}$ is given in Appendix \ref{appendixQ}.
We also note that $\mathcal Q^{\rm id}$ is the idealized unitary based on ideal rotations. When non-ideal, the action of $\mathcal{Q}$ no longer remains in the two-dimensional subspace of $\ket{{\chi^{\rm id}}}$ and $\mathcal{V}\ket{{\chi^{\rm id}}}$.
We discuss the effect of erroneous applications of the gates in Sec.~\ref{sectionErrors}.

\subsection{Adding a control mode}

In the qubit algorithm, for phase estimation we apply the controlled version $\mathcal Q_c$ of an operator $\mathcal{Q}$ that transforms as
\begin{equation}
\mathcal Q_c \ket j \ket \psi = \ket j \mathcal Q^j \ket \psi, 
\end{equation}
where $\ket j$ is a label state comprised of multiple qubits. Such an operation can be built up from the controlled unitary 
\begin{equation}
\ket 0\bra 0 \otimes \mathbbm I + \ket 1\bra 1 \otimes \mathcal Q, 
\end{equation}
where we apply $\mathcal Q$ if a single control qubit is in the state $\ket 1$ and do nothing otherwise. 

In the CV setting, we modify this approach by replacing a register of control qubits by a single resource mode denoted by $\phi$, which is prepared in a position eigenstate $\ket {x}_{q_\phi}$, see e.g. Ref.~\cite{Liu2016}.
We attach $\phi$ to the current three modes by performing a four mode interaction that can be seen as a controlled version of $\mathcal{Q}^{\rm id}$. In particular,
$\phi$ is attached via the operator $\hat p_\phi$, leading to the ideal phase estimation operator
\begin{eqnarray}
\mathcal Q^{\rm id}_c
&=& e^{-i 1/\sqrt{f(\hat q_1)}\otimes \hat p_2 \otimes \hat p_3 \otimes \hat p_\phi}   (\times {\rm more\ terms}).
\end{eqnarray}
The full expression for $\mathcal Q^{\rm id}_c$ is given in Eq.~\eqref{Eq:BigQcid}. This interaction requires $\mathcal G_c$ as the controlled version of $\mathcal G$. More details are given in Appendix \ref{appendixQ}. We note that the addition of control through $\mathcal Q^{\rm id}_c$ must be efficient for speedups in CV QMC, see Sec.~\ref{sectionErrors} for further discussion.

\subsection{Phase estimation}
\label{sectionPhase}

The operator $\mathcal Q_c$ is used to perform phase estimation to extract the eigenvalues of
$\mathcal Q$, which are related to the desired integral. 
We briefly show this schematically, before providing a more precise analysis. 
Let $\ket {\psi_+}$ be the eigenstate of $\mathcal Q^{\rm id}$ corresponding to the eigenvalue $e^{i\theta}$, with $\theta$ as given by Eq.~(\ref{eqTheta}). Then, with the phase estimation mode in the initial state $\ket 0_{q_\phi}$, 
\begin{equation}
\mathcal Q_c^{\rm id} \ket {\psi_+} \ket 0_{q_\phi} =  \ket {\psi_+} e^{i \theta \hat p_\phi} \ket 0_{q_\phi} =  \ket {\psi_+} \ket \theta_{q_\phi},
\end{equation}
where the phase estimation mode is shifted in position by an amount equal to $\theta$. 
A position measurement of the $\ket \theta_{q_\phi}$ mode then obtains the result. 

Performing phase estimation with a single infinitely squeezed mode allows for $\theta$ to be measured exactly with a single measurement. However, this is clearly unphysical, and we now perform an analysis of phase estimation using a finitely-squeezed state centered around $0$, i.e., the state $\ket{G_{0,s}}$.
Let again $\ket {\psi_+} $ be the eigenstate of $\mathcal Q^{\rm id}$ with eigenvalue $e^{i\theta}$.  
Define $\epsilon_\theta^{\rm target}>0$ to be the desired final accuracy for the $\theta$ value.
Note that the final error for approximating $\mathcal{I}$ is then also $\mathcal{O}(\epsilon^{\rm target}_\theta)$~\cite{Rebentrost2018finance}.
Let $M$ be an integer. We apply $\mathcal Q^{\rm id}_c$ for $M$ times, leading to  
\begin{eqnarray}\label{eqPhaseEstimationState}
\left(\mathcal Q^{\rm id}_c \right )^M \ket {\psi_+} \ket {G_{0,s}} &=&  \ket {\psi_+} e^{-i M \theta \hat p_\phi} \ket {G_{0,s}}\\
&=&   \frac{\ket {\psi_+}}{\sqrt{s}{\pi^{1/4}}} \int dx e^{-\frac{x^2}{2s^2}} \ket {x+ M \theta}_{q_\phi}. \nonumber
\end{eqnarray}
Measuring the position of the resource mode obtains a result sampled from the error-free probability distribution
\begin{eqnarray}
P_\theta(q) =  \frac{1}{s \sqrt{\pi}}  e^{-\frac{(M\theta-q)^2}{s^2}}.
\end{eqnarray}
See Appendix \ref{appendixPE} for the expression for $P_\theta(q)$ including the error 
due to the erroneous simulation of $\mathcal Q_c$, which is enumerated by $\epsilon_{Q}$. 

Let the samples obtained from independent runs be given by $Y_j$. 
The success probability of a single measurement $Y_j/M$ being inside a range $\epsilon^{\rm target}_\theta$ around the expectation value $\theta$, i.e., $\vert Y_j/M - \theta \vert \leq \epsilon^{\rm target}_\theta$, is given by
\begin{equation}\label{eqSuccessProb}
p_{\rm success} = {\rm erf}\left( \frac{ M\epsilon^{\rm target}_\theta}{\sqrt{(M\epsilon_Q)^2 + s^2}}\right),
\end{equation}
see Appendix \ref{appendixPE}, Eq.~(\ref{eqSuccessProb}). 
Using $M \geq 1/\epsilon^{\rm target}_\theta$ \cite{Xu2018}, the vacuum squeezing $s=1/\sqrt{2}$, and no gate errors $\epsilon_Q=0$, we can lower bound the success probability to be $p_{\rm success} \geq {\rm erf}(\sqrt 2) > 0.95$. In the presence of gate errors  $\epsilon_{Q}$, if we pick $\epsilon_{Q} = \epsilon^{\rm target}_\theta$ and vacuum squeezing again, we obtain a lower bound of 
$p_{\rm success} \geq {\rm erf}(\sqrt{2/3}) >  0.75$.

A single-shot success probability greater than $1/2$ is a requirement for boosting the success probability via multiple independent runs.  We repeat the measurement $L\geq 1$ times and take the median of the obtained values $Y_j/M$.
Let the desired success probability for $\vert {\rm Median}(Y_j/M) - \theta \vert \leq \epsilon^{\rm target}_\theta$, or ``confidence", be given by $c$. It can be achieved by using $L \leq \vert \log(1-c) \vert/ \vert \log(2\sqrt{p_{\rm success}(1-p_{\rm success})}) \vert$ repetitions \cite{Nagaj2009,Rebentrost2018finance}. Concretely, if we have $p_{\rm success}={\rm erf}(\sqrt{2/3})$ and would like to boost it to $c=0.995$, then $L \approx 37$ will be sufficient.

Furthermore, we can leverage squeezing in the phase estimation mode as a resource. Indeed, choose $s$ smaller than the vacuum squeezing, $s < 1/\sqrt{2}$, $\epsilon_{Q} = \epsilon^{\rm target}_\theta$, but leave $M$, the number of required applications of $\mathcal Q_c$, as a variable. To achieve the same success probability ${\rm erf}(\sqrt{2/3})$, the argument of the error function has to be the same which leads to the 
relation $(M \epsilon^{\rm target}_\theta)^2 = 2s^2$. Thus we can take $M \geq \sqrt{2} s /\epsilon^{\rm target}_\theta$, which lowers the requirement for $M$ at the cost of more squeezing $\sqrt{2}s<1$.
The feasibility of implementing corresponding squeezing factors will have to be determined for each application and for different experimental setups. 

\subsection{Query complexity speedup}\label{Sec:Speedup}

In QMC, we conventionally use the number of applications of the relevant unitary to consider the speedup. Here, we consider applications of $\mathcal{K}$, since it is the unitary used to encode the integral from the first stage of the algorithm. In physical implementations, $\mathcal{Q}_{c}$ and its composing elements have a runtime which must be accounted for if any real speedup is to occur. Runtimes are discussed in the following section on efficiency and errors.

Classical algorithms typically require
$N_{C}=\Ord{1/\epsilon_\theta^2}$ evaluations of the integrand, see Eq.~(\ref{eqCheb}).
In the quantum algorithm, each application of $\mathcal{Q}_{c}$ involves a constant number of applications of $\mathcal K$, see Eq.~(\ref{Eq:Q}).
The total number of applications of 
$\mathcal K$ is thus
\begin{equation}\label{eqNq}
N_{Q} = \Ord{M \times L}.
\end{equation}
To obtain a quantum speedup,  we take $M \geq \sqrt 2 s/\epsilon_{\theta}^{\rm target}$ and
the gate error $\epsilon_Q = \epsilon_\theta^{\rm target}$. We also take
$L$ to be constant (e.g. $\approx 37$), as discussed.
This means that 
\begin{equation}\label{Eq:Speedup}
N_{Q} = \Ord{\frac{s}{\epsilon_\theta^{\rm target}}}.
\end{equation}
Thus, a quadratic speedup is obtained over the classical runtime. The achieved error is $\epsilon_\theta^{\rm target}$ and the confidence is 
high, say $99.5\%$.
Surprisingly, further speedups may be possible by concurrently decreasing $M$ and decreasing $s$. We note, however, general lower bounds for the search and amplitude amplification problems \cite{Bennett1997}.

\subsection{Gate complexity}

We now summarize the gate complexity of the algorithm.
The classical complexity is $\tOrd{1/\left( \epsilon_\theta^{\rm target}\right)^2}$ for the common situation when the integrand can be evaluated classically in $\Ord{{\rm poly} \log 1/\epsilon}$. Here, $\tOrd{\cdot}$ omits polylogarithmic factors. 
For the quantum algorithm, let a single application of $\mathcal Q$ be achieved to error $\epsilon_Q$ at a runtime cost $T_Q\left(\epsilon_Q\right)$. Using the requirement $\epsilon_Q =  \epsilon_\theta^{\rm target}$ obtains a runtime $T_Q\left( \epsilon_\theta^{\rm target} \right)$.
Together with the number of calls to the unitary $\mathcal Q$, the total gate complexity scales as
\begin{equation}
N_q T_Q\left ( \epsilon_\theta^{\rm target} \right) = \Ord{ \frac{s T_Q\left( \epsilon_\theta^{\rm target} \right)}{ \epsilon_\theta^{\rm target}}}.
\end{equation}
Thus in terms of the gate complexity, the possibility of a speedup crucially depends on $T_Q( \epsilon_\theta^{\rm target} )$.
For a quadratic speedup we require
\begin{equation}
T_Q\left( \epsilon\right) = \Ord{ {\rm poly} \log \left(1/ \epsilon\right) },
\end{equation}
which then results in 
\begin{equation}
N_q T_Q\left ( \epsilon_\theta^{\rm target} \right) = \tOrd{\frac{s}{\epsilon_\theta^{\rm target} }}.
\end{equation}
We can still achieve speedups below quadratic if we assume a $0 < \delta<1$ and
\begin{equation}
T_Q( \epsilon ) = \Ord{1/ \epsilon^\delta }.
\end{equation}
which then results in 
\begin{equation}
N_q T_Q \left( \epsilon_\theta^{\rm target} \right) = \Ord{\frac{s}{\left(\epsilon_\theta^{\rm target}\right)^{1+\delta} }}.
\end{equation}
With these assumptions on the gate complexity of $\mathcal Q$, the overall quantum gate complexity consequently scales better than the classical complexity.
The next sections discusses the assumptions for such a runtime on the individual quantum operations and methods to achieve such a runtime. 
\section{Errors and gate complexity}
\label{sectionErrors}

Here we account for the errors that arise due to various approximations in a physical implementation of CV QMC. These errors can be split into different categories: (A) due to preparing the first mode, (B) due to finite squeezing, (C) from the polynomial approximation of $f(x)$ and (D) through implementing various gates/unitaries. We also discuss here the effect of gate runtimes on any speedups through CV QMC by using as a proxy the number of required Gaussian unitaries and cubic phase gates. The referenced appendices support this section.

\subsection{Errors in preparing $p(x)$}

The first mode is prepared according to $\sqrt{p(x)}$ by applying the unitary $\mathcal{G}$ to the vacuum. We suppose that $\mathcal G$ can be applied to accuracy $\epsilon_G$, and moreover that this can be achieved using $T_G$ continuous-variable Gaussian and cubic phase gates.
For a quadratic speedup in CV QMC we require that
\begin{equation}
T_G = \Ord{{\rm poly} \log 1/\epsilon_G}.
\end{equation}
We further assume that controlling $\mathcal G$ adds at most a logarithmic overhead to the runtime. For sub-quadratic speedups, these requirement can be relaxed. 
\subsection{Squeezing errors}

Generally there will be a maximum required squeezing and a maximum achievable squeezing in any given mode. Assume that there exists a squeezing factor $r_{\Delta x'}$ such that larger squeezing leads to computationally equivalent results. Equivalently, there is a scale $\Delta x'$ such that two CV position states $\ket x$ and $\ket {x+\Delta x'}$ are computationally equivalent. 
Moreover, there is a physically achievable squeezing factor given by $r_{\max}$. 
The interesting case is $r_{\max} \leq r_{\Delta x'}$, when the achievable squeezing is below the computationally required squeezing, requiring an error analysis.

The present CVMC algorithm and other continuous-variable algorithms schematically use the infinite squeezing gate
\begin{equation}
S(\infty),
\end{equation}
here for the gate $\mathcal G\otimes S(\infty) \otimes S(\infty)$ in Eq.~(\ref{eqOpK}).
By assumption, \textit{without} changing the effect of the computation, we can replace
\begin{equation}
S(\infty) \to S(r_{\Delta x'}),
\end{equation}
where $r_{\Delta x'}$ is the squeezing factor corresponding to $\Delta x'$ according to Eq.~(\ref{eqSqueezingFactorsRelation}).
In a physical implementation, squeezing factors of $r_{\Delta x'}$ may not be reached. Instead, we reach $r_{\max}$.
The gate error due to only reaching $r_{\max}$ can be quantified via
\begin{eqnarray} \label{eqSError}
\epsilon_S := \left \Vert S(r_{\max}) - S(r_{\Delta x'})\right \Vert &=& 
\Ord{ \vert r_{\Delta x'}-r_{\max}\vert}.
\end{eqnarray}

There are four modes in CV QMC which require various degrees of squeezing: (1) the first mode may require squeezing for state preparation, (2) the second and third mode require squeezing for performing the controlled rotation and also to implement the projector $\mathcal P$. Finally, (3) the fourth mode is squeezed according to $s$ for phase estimation. Case (1) is accounted for in $\epsilon_{{G}}$, case (2) is handled by 
Eq.~(\ref{eqSError}) and also in Appendix~\ref{App:FiniteSqueezing}, while case (3) determines the final error $\epsilon_{\theta}$ in estimating the integral. 

\subsection{Polynomial approximation errors}

\begin{figure}
\includegraphics[width=\columnwidth]{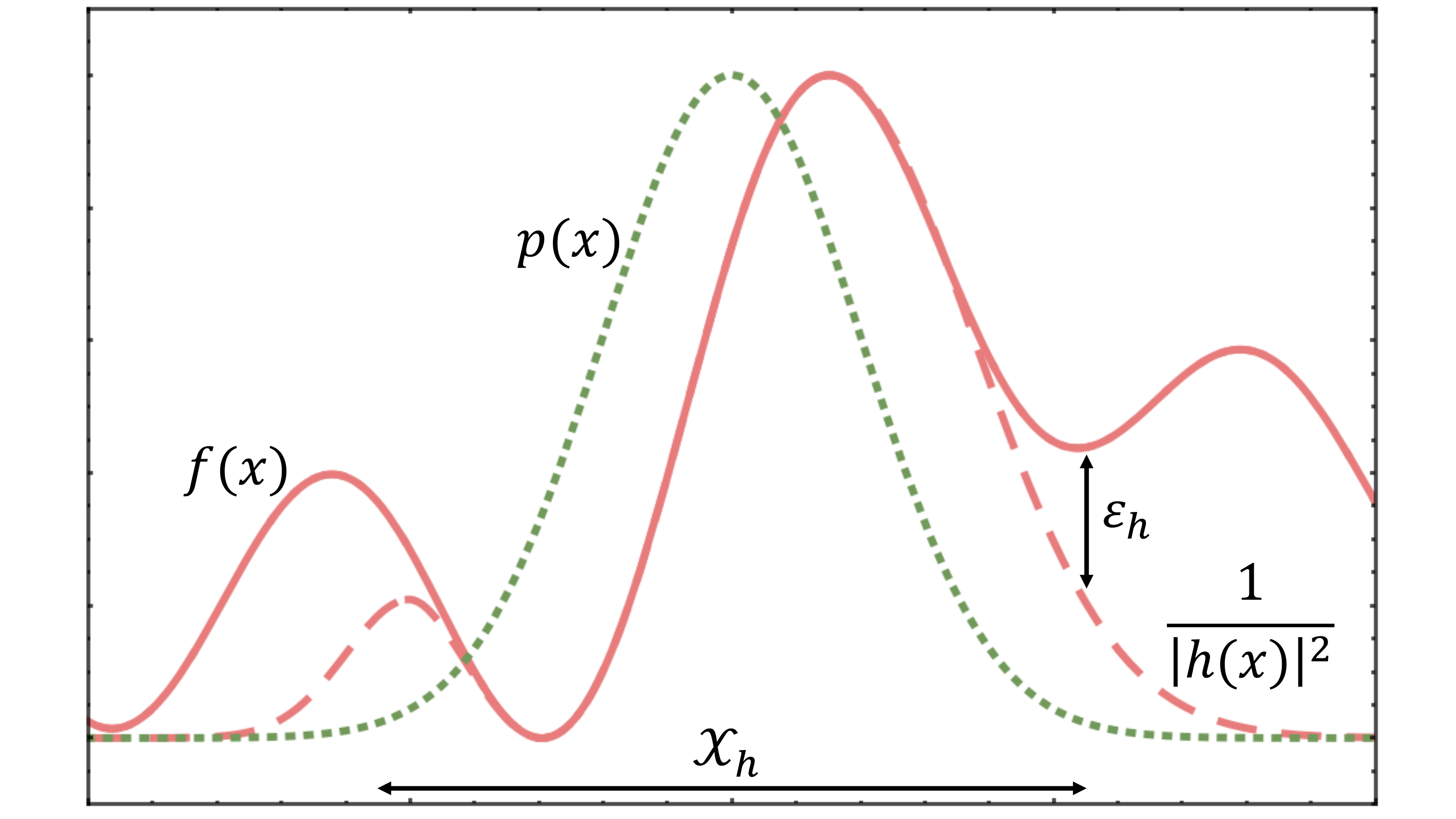}
	\caption{Example polynomial approximation of $f(x)$. Here, $p(x)$ (dotted green line) is a Gaussian distribution and $f(x)$ is an arbitrary function (solid red line). The set $\mathcal{X}_{h}$ selects the non-trivial region of the integral, and we find a polynomial $h(x)$ such that $1/|h(x)|^{2}$ (dashed red line) approximates $f(x)$ within $\mathcal{X}_{h}$ up to some error $\epsilon_{h}$.}
\label{Fig:PolyApprox}
\end{figure}

Recall that $0 \leq f(x) \leq 1$ and suppose that there exists a polynomial $h(x)$ such that $1/|h(x)|^{2}$ approximates $f(x)$ in the following way.
Define a compact set $\mathcal X_h$ which 
 denotes the non-trivial region of the integral. First, 
we require that $h(x)$ satisfies the point-wise error condition
\begin{equation}\label{eqErrorPointwise}
\left \vert f(x) - \frac{1}{\vert h(x) \vert^2} \right \vert \leq \epsilon_h,
\end{equation}
for all $x \in \mathcal X_h$ where $\epsilon_h>0$, see Fig.~\ref{Fig:PolyApprox}.
Outside the set, we require that
\begin{equation}\label{Eq:TrivialRegion}
\int_{ \bar{\mathcal X_h}} dx p(x) \left \vert \frac{1}{\vert  h(x) \vert^2}  - f(x) \right \vert \leq \eta
\end{equation}
for a small $\eta > 0$, where $\bar{\mathcal X_h}$ is the complement.
The total error $ \left \vert \int dx \frac{p(x)}{\vert  h(x) \vert^2}  - \int dx f(x)p(x) \right \vert$ is then $\mathcal{O}(\epsilon_{h} + \eta)$, see Appendix \ref{appendixErrorPolynomial}. Thus, under the assumptions and $\eta=\Ord{\epsilon_h}$, the polynomial approximation error is no larger than $\Ord{\epsilon_h}$. 

\subsection{Controlled rotation error}

Regarding the gate $\mathcal H$, 
the pointwise error property Eq.~(\ref{eqErrorPointwise}) leads to the gate error 
\begin{equation}
\Vert \mathcal H^{\rm id} - \mathcal H^{\rm impl} \Vert =\Ord{\epsilon_h},
\end{equation}
when the polynomial argument is inside $\mathcal X_h$. Appendix~\ref{appendixCVBasics} defines the operator norm used in this work.
This is shown via 
$\left \Vert \mathcal H^{\rm id}  - \mathcal H^{\rm impl} \right \Vert = \Ord{\left \vert h(x) - \frac{1}{\sqrt{f(x)} } \right \vert }= \Ord{f(x)^3 \epsilon_h} = \Ord{ \epsilon_h}$, using $f(x)\leq 1$.

In addition, the gate $\mathcal H$ can in principle decomposed into elementary Gaussian operations and cubic phase gates, denoted by an operator $\mathcal H^{\rm dec}$.
The cost shall be given by $T_H(d_{h},\epsilon_{h}^{\rm dec})$ to error $\epsilon^{\rm dec}_h := \Vert \mathcal H^{\rm impl} - \mathcal H^{\rm dec}\Vert$, where $d_{h}$ is the polynomial degree of $h(x)$. 

Regarding this cost, if the decomposition requires a number of gates $T_H=1/\epsilon_{h}^{\rm dec}$ then possibilities of a quantum speedup are lost. Such runtime costs appear for example when lowest-order Suzuki-Trotter methods are used \cite{Sefi2011}. With higher-order Suzuki -Trotter methods, one can in principle achieve a runtime of $T_H=\Ord{1/(\epsilon_{h}^{\rm dec})^\delta}$ with a constant $0<\delta <1$ \cite{Childs2017} that can be made arbitrarily close to $0$. Such methods are expected to translate to the CV context, but a proper analysis will be left for future work. Additionally, we note that exponentially precise Hamiltonian simulation methods exist for qubits \cite{berry2017exponential,Gilyen2018}, which may also be translated into the CV framework. In such cases, we may even achieve $T_H=\Ord{\log 1/\epsilon_{h}^{\rm dec}}$.
The dependence of $T_H$ on $d_h$ is usually $\Ord{2^{d_h}}$ with a potential $\Ord{{d_h}}$ discussed in Appendix \ref{appendixDecomposition}.
In summary, $\mathcal H$ can be implemented to accuracy $\epsilon_H := \Vert \mathcal H^{\rm id}  - \mathcal H^{\rm dec} \Vert = \Ord{\epsilon_h + \epsilon^{\rm dec}_h}$ in runtime $T_H$.
We further assume that controlling $\mathcal H$ adds at most a logarithmic overhead to the runtime.

\subsection{Reflections}
The reflection gates $\mathcal V$ and $\mathcal Z$, defined in Eqns.~(\ref{eqV}) and (\ref{eqZ}), respectively, are also performed to an error. Possible methods for implementing them are shown in Appendix \ref{SecReflec}.
The gate $\mathcal Z$ is implemented with squeezing and application of the PBL gate, see Appendix \ref{appendixZ}.
The accuracy $\epsilon_Z$ depends on the parameter $\Delta x$ of the PBL gate and the squeezing error, obtaining $\epsilon_Z = \Ord{\Delta x^2 + \epsilon_S}$. 
 For the runtime, we obtain a constant number of applications of the Pati, Braunstein, and Lloyd (PBL) gate. 
This suggests that potentially 
$T_Z = \Ord{{\rm poly} \log 1/\epsilon_Z}$ but further research has to be devoted to the efficient implementation of the PBL gate. 

The gate $\mathcal V$ is  be implemented via squeezing, displacements, and the PBL gate, see Appendix \ref{appendixV}
 to accuracy $\epsilon_V = \Ord{\Delta x^2 + \epsilon_S}$. The runtime potentially is $T_V=\Ord{{\rm poly} \log 1/\epsilon_Z}$ since a constant number of PBL gates are required.

\subsection{The operator $\mathcal Q$}
\label{sectionErrorQ}
The first stage of the algorithm is to apply $\mathcal{K}$ in Eq.~\eqref{eqOpK}, while the second stage involves multiple repetitions of $\mathcal{Q}$ using the reflections $\mathcal{V}$ and $\mathcal{Z}$ for phase estimation, along with $\mathcal{K}$ (see Eq.~\eqref{Eq:Q}). Each of these gates has an error and corresponding runtime.
The operator $\mathcal K$ is in turn composed of state preparation $\mathcal{G}$, squeezing of the ancilla modes, and the controlled rotation $\mathcal{H}$. 
The sequence of operations is shown in Eq.~(\ref{Eq:BigQid}).
Thus, implementing $\mathcal Q$ achieves an accuracy 
\begin{equation} \label{eqErrorQ}
\epsilon_Q = \Ord{\epsilon_G + \epsilon_S + \epsilon_H + \epsilon_Z + \epsilon_V},
\end{equation}
and
requires runtime
\begin{eqnarray}
T_Q(\epsilon_G, \epsilon_H, \epsilon_Z, \epsilon_V) &=& 4T_G(\epsilon_G) + 4T_H( \epsilon_H) \nonumber \\
&&+ 2 T_Z( \epsilon_Z) + 2 T_V( \epsilon_V). 
\end{eqnarray}
To simplify the analysis, we take the allowable error for the individual operations proportional to a small constant times $\epsilon_Q$, i.e., 
we take the required errors to be $\epsilon_i = \Ord{\epsilon_Q}$ such that they add up to $\epsilon_Q$. 
For the runtime, if each of the individual $T_i( \epsilon_i) = \Ord{{\rm poly} \log 1/\epsilon_i}$ either by assumption or by the employed methods, and because of Eq.~(\ref{eqErrorQ}), we can find the overall bound $T_Q(\epsilon_Q) = \Ord{{\rm poly} \log 1/\epsilon_Q}$. Otherwise, if one of the individual $T_i( \epsilon_i) = \Ord{1/\epsilon_i^\delta}$ with $0<\delta<1$ we  find the overall bound $T_Q(\epsilon_Q) = \Ord{ 1/\epsilon_Q^\delta}$.

We further assume that controlling $\mathcal Q$ via controlling all the gates adds at most a logarithmic overhead to the runtime. 

\section{Multidimensional integration}
\label{secMulti}

The generalization of this algorithm for $n$-dimensional integration, i.e., Eq.~\eqref{eqExpValMain}, is straightforward. We begin by preparing $n$ modes according to $p(\vec{x})$ by
applying the operator $\mathcal G$ 
to obtain 
\begin{equation}
\mathcal{G} \ket{\rm vac}^{\otimes n} = \int d \vec{x} \,\, \sqrt{p(\vec{x})} \ket{\vec{x}}_{q},
\end{equation}
where $\ket{\vec{x}}_{q}$ is the product of position eigenstates corresponding to $\vec{x}$.
Let  $h(\vec x) = h(x_1,\dots,x_n)$ be a polynomial that suitably approximates $ f(\vec x)$ via  $\frac{1}{\vert h(\vec x)\vert^2}$.
Define the gate acting on the $n$ modes plus two more ancilla modes as
\begin{equation}\label{eqGateHMult}
\mathcal H := e^{-i  h(\hat q_1,\dots,\hat q_n)\otimes \hat p_{n+1} \otimes \hat p_{n+2}}.
\end{equation}
Applying the gate to $\mathcal{G}\ket{\rm vac}^{\otimes n}\ket{0}_{q_{n+1}}\ket{0}_{q_{n+2}}$
gives
\begin{equation}
\int d \vec{x} \,\, \sqrt{p(\vec{x})} \ket{\vec{x}}_{q} \int d p' \,\, \ket{p'}_{p_{n+1}}\ket{h(\vec{x}) p'}_{q_{n+2}}.
\end{equation}

The remaining analysis is analogous to the single variable case discussed above. 
Define the operator 
\begin{equation}
\mathcal{M} :=  \mathbbm I^{\otimes n} \otimes \ket{0}_{q_{n+1}} \bra{0}_{q_{n+1}} \otimes \ket{x_{\rm off}}_{q_{n+2}} \bra{x_{\rm off}}_{q_{n+2}},
\end{equation}
i.e., the extension to $n+2$ modes of the operator $\mathcal{M}$ in Eq.~\eqref{Eq:M} used to extract the integral. 
Then the expectation value of this measurement on the state above is given by 
\begin{equation}
\langle \mathcal{M} \rangle = \frac{1}{4\pi} \int d\vec x  \,\, \frac{p(\vec x)}{\vert  h(\vec x) \vert^2} \propto \mathcal{I}.
\end{equation}
The error analysis and extension to finite squeezing is analogous to the one-dimensional case. Phase estimation proceeds similarly by adding another mode and using the phase estimation operator
\begin{equation}
\mathcal{Q}_{c}^{\rm id} = e^{-i h(\hat q_1,\dots,\hat q_n)\otimes \hat p_{n+1} \otimes \hat p_{n+2} \otimes \hat p_\phi} \,\, (\times \rm more \,\, terms),
\end{equation}
which is the generalization of Eq.~\eqref{Eq:BigQcid}.

\section{Numerics}\label{Sec:Numerics}

As discussed previously, one needs $n+3$ optical modes with appropriate squeezing
in order to evaluate an $n$ dimensional integral using the CV QMC algorithm. There
are two main ingredients: (i) encoding of the integrand  and (ii) amplitude estimation using amplitude amplification and phase estimation. The first $n+2$ modes encode the
underlying integrand while an ancilla mode is added for amplitude estimation, so that repeated measurements of its position lead to the expected value of the integral. While the encoding of the
integrand and amplitude amplification can be hard to simulate, we can showcase the quadratic speedup of one mode phase estimation, where the integral-dependent phase $\theta$ is predetermined classically, following the approach taken in
Ref. \cite{Rebentrost2018finance}.

In this section, we consider a simple example with 
$f(x) = 1/(1+x^2)^2$ subject to $p(x)$ taken to be a Gaussian probability distribution $G_{x_{0},s}$ with $x_{0} = 0$ and $\sigma = 1/2$. With these choices, Eq. (\ref{eqExpValMain}) can be
integrated analytically giving $\mathcal I \approx 0.74$. We take $\theta \approx 0.98$ to
be the predetermined phase for the single mode phase estimation on the ancilla
mode, i.e., the solution to
\begin{equation}
\theta = 2 \arccos\left(1 - \frac{\mathcal{I}}{2\pi}\right).
\end{equation}
Single mode phase estimation is carried out using the Strawberry Fields software suite~\cite{Killoran2018}. We also estimate $\mathcal I$ using the standard classical
Monte Carlo integration techniques and, at the end, compare the scaling of the
errors with number of MC steps used in the two algorithms.

Let $\hat \theta$ and $\theta$ (which in the particular example considered here
is 0.98) be the estimated and predetermined values of the phases, respectively.
We define the corresponding estimation error as the fractional difference between
$\hat \theta$ and $\theta$:
\begin{equation}\label{error}
  {\rm Error} := |\hat\theta-\theta|/\theta.
\end{equation}
In the following, the subscripts $Q$ and $C$ denote the quantum and classical
estimations. The fractional error defined above follows a power-law behavior
with the number of MC steps as follows:
\begin{equation}\label{powerlaw}
  {\rm Error} = b~N^\zeta,
\end{equation}
where $N$ is the number of MC steps and $\zeta$ denotes the scaling exponent.
For the standard classical MC approach the scaling exponent is $\zeta_{C} =
-\frac{1}{2}$.

\begin{figure}
\includegraphics[width=\columnwidth]{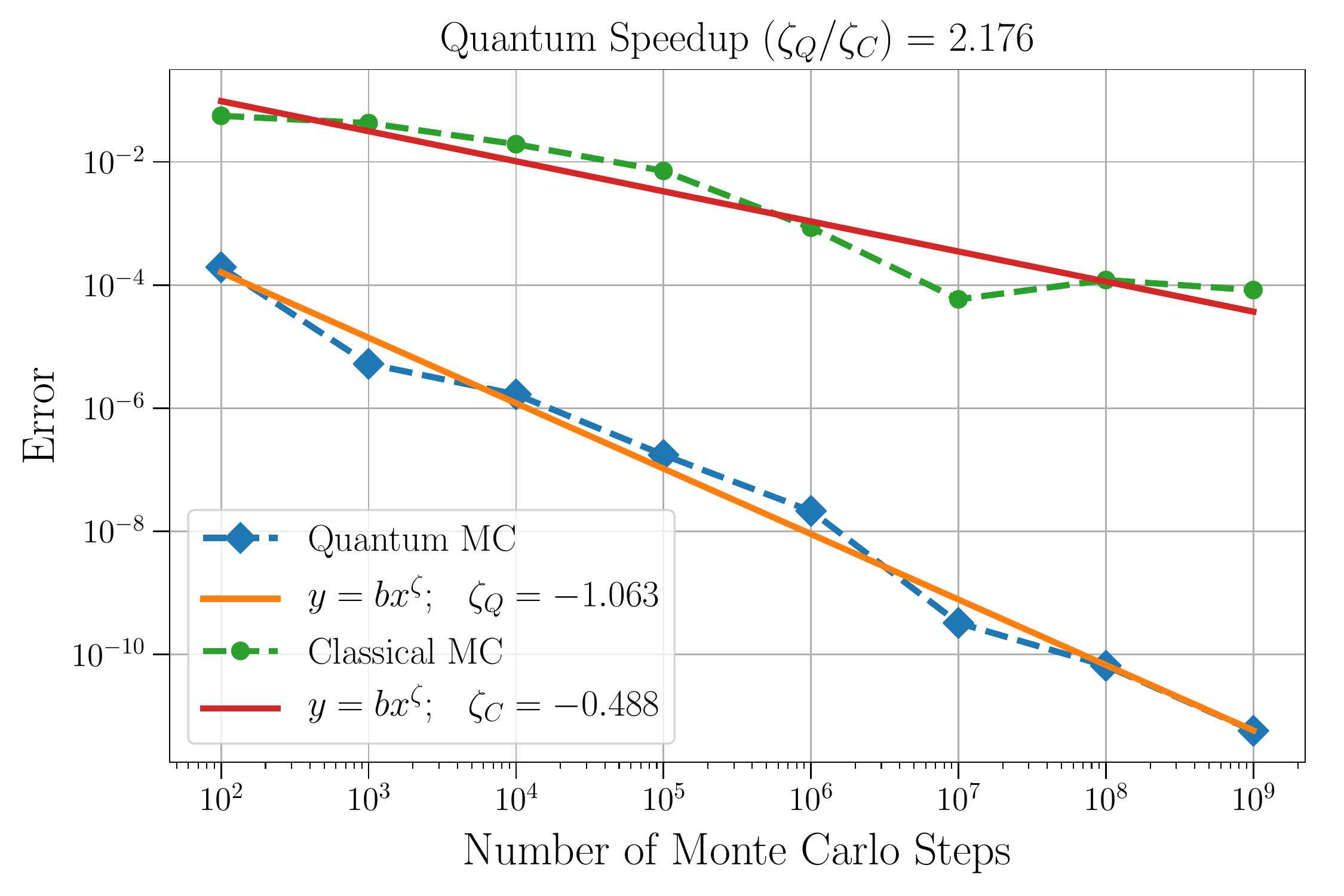}
	\caption{Comparison of classical and quantum MC. Fractional error (defined in Eq. \eqref{error}) in the classical and
quantum estimations are plotted against the number of MC steps ($N$) and fitted to a
power law function. The squeezing parameter is fixed to $r=10$. It is evident
that quantum error scales approximately quadratically faster than the classical
counterpart, i.e. $\zeta_{Q}/\zeta_{C}\approx 2$.
}
\label{fignumerics}
\end{figure}

Figure \ref{fignumerics} shows the comparison between the classical and quantum
error scalings. The left panel shows the behavior of the error as a function of
MC steps. The dotted-dashed curves show the data points and the solid lines are
fits to the power law behavior of Eq. \eqref{powerlaw}. For the plot shown in the left panel 
we fixed the squeezing parameter to be $r=10$. The plots illustrate that for the quantum algorithm $\zeta_{Q} \approx -1.063$, giving an
advantage over the classical method by a factor of $\zeta_{Q}/\zeta_{C}\approx 2$. This indicates an approximately quadratic speedup in the error scaling in
terms of the number of MC steps. For the quantum case we consider $N_{Q} = ML$, where we have fixed $L=100$ and varied $M$. 

\section{Discussion and Conclusion}
\label{sectionDiscussion}

We have discussed a CV quantum algorithm for MC integration. 
To summarize, the main assumptions for the algorithm to work are the following. 
First, there are certain direct requirements on the integrand.
\begin{enumerate*}[label=(\roman*)] 
\item The function $f(\vec x)$ is bounded as $0 \leq f(\vec x) \leq 1$ for all $\vec x$. 
This also means that the desired integral is $\int p(\vec x) f(\vec x) d\vec x \leq 1$. An extension to more general functions was given in the qubit context in Ref.~\cite{Montanaro2015} and will be subject of future work. 
\item There exists a polynomial $h(\vec x)$ which relates to the function $f(\vec x)$ via $f(\vec x) = 1/ \vert h(\vec x)\vert^2$ with point-wise error at most $\epsilon_h$ on a compact set. Outside of the set, the integral is vanishingly small.
\end{enumerate*}
Next, there are requirements on implementations of the algorithm to provide a speedup over classical methods.
\begin{enumerate*}[label=(\roman*)]
\setcounter{enumi}{2}
\item There exists a unitary $\mathcal G$ to efficiently prepare a quantum state which encodes  $\sqrt{p(\vec x)}$ in its amplitudes.
\item We assume that an efficient continuous-variable gate sequence related to the polynomial function $h(\vec x)$ can be constructed.
\item Lastly, a reflection around the computational initial state and the state defining the projective measurement can be efficiently implemented, i.e., the gates $\mathcal{V}$ and $\mathcal{Z}$. 
\end{enumerate*} 

Until now, only the Grover search problem has been discussed in the CV framework \cite{Pati2000}, and the generalization to amplitude estimation and MC simulations was not provided in the literature.
The algorithm presented here can potentially achieve quadratic speedups in estimating integrals on a continuous-variable quantum computer. Moreover, the CV setting naturally accommodates the task of multidimensional integration and requires a fixed number of modes, this contrasts with the qubit setting of QMC which requires discretization and a number of qubits that increases with the desired accuracy.

For the CV amplitude estimation, we have discussed the important role of the vacuum reflection and
high-quality gate decompositions. We have constructed an implementation of the reflection by expressing it in terms of a gate introduced previously \cite{Pati2000}. However, simpler implementations may be possible to achieve the desired relative phase of the zero-photon state in analogy to the qubit implementations. 
Gate decompositions are an important ingredient of the present work. For amplitude estimation to provide speedups, such gate decompositions have to be efficient with small errors. Higher-order Suzuki-Trotter expansions can in principle achieve 
a $1/\epsilon^\delta$ runtime dependency in the error $\epsilon$, where $\delta$ is a constant that can be made arbitrarily small. While there are many studies on gate decompositions for qubits, the detailed study of such expansions and even exponentially precise methods is still in its relative infancy in the CV setting.

The applications of Monte Carlo integration are manifold, for example in mathematical finance (the pricing problem \cite{Rebentrost2018finance}) and machine learning (Markov chain sampling \cite{Szegedy2004}).
Future work will include steps toward concrete implementations of the algorithm presented here on realistic photonic hardware. It will also be of value to investigate quantum generalizations and applications of the two other main areas of MC methods: optimization and sampling.

\acknowledgements
We acknowledge Juan Miguel Arrazola, Timjan Kalajdzievski, and Krishna Kumar Sabapathy for insightful discussions.

\bibliographystyle{apsrev}
\bibliography{Qmc}

\onecolumngrid

\appendix

\section{CV basics}
\label{appendixCVBasics}

This section reviews the CV framework. The continuous-variable quantum computational model uses the quantized light field for computation. The quantization can be described by the canonical operators $\hat q$ and $\hat p$ which satisfy the commutation relation $[\hat q, \hat p] = 2i$, and are called position and momentum quadratures, respectively. Alternatively one can define ladder operators $\hat a^\dagger$ and $\hat a$, which describe the creation and annihilation of energy quanta.
They are related to the quadratures via
\begin{eqnarray}
\hat a &=& \frac{1}{2} (\hat q+i \hat p),\\
\hat a^\dagger &=& \frac{1}{2} (\hat q-i \hat p),
\end{eqnarray}
and, conversely, 
\begin{eqnarray}
\hat q &=& (\hat a^\dagger + \hat a),\\
\hat p &=& i (\hat a^\dagger - \hat a).
\end{eqnarray}
The (unnormalized) eigenstates of the position quadrature $\hat q$ are given by the states $\ket x_q $ for which,
\begin{equation}
\hat q \ket x_q  = x \ket x_q.
\end{equation}
The state $\ket x_q $ denotes the mode in a position eigenket corresponding to the position $x$.
For the momentum quadrature we have equivalently, 
\begin{equation} 
\hat p \ket x_p = x \ket x_p,
\end{equation}
where $\ket x_p,$ denotes the mode in a momentum eigenket corresponding to the momentum $x$.
The $\hat q$ and $\hat p$ eigenstates can be related to each other via a Fourier representation. 
The relations are given by
\begin{eqnarray}
\ket {q'}_q &=& \frac{1}{2\sqrt{\pi}} \int dp' e^{-i q' p'/2} \ket {p'}_p, \\
\ket {p'}_p &=& \frac{1}{2\sqrt{\pi}} \int dq' e^{i q' p'/2} \ket {q'}_q.
\end{eqnarray}
Sometimes we write in shorthand $\ket {q'}_q \to \ket{ q'}$ and $\ket {p'}_p \to \ket {p'}$ if the context is clear. 
For the inner product it holds
\begin{eqnarray}
\braket{p'}{q'}_{p,q} = \frac{e^{-i q'p'/2}}{2\sqrt{\pi}}.
\end{eqnarray}
A continuous-variable quantum state is fully described by its expansion into positions eigenvectors (or, equivalently, momentum eigenvectors.) Given a state $\ket \psi$ it can be expanded into $\int dq \,\, \psi(q) \ket q$, with the expansion coefficients given by the  wavefunction in position space $\psi(q)$. To translate into momentum space, we obtain
\begin{equation}
\int dq \,\, \psi(q) \ket q =  \frac{1}{2\sqrt{\pi}}  \int dq \,\, \psi(q) \int dp e^{-i q p/2} \ket p =: \int dp \,\, \psi^{\rm FT}(p) \ket p.
\end{equation}
with the wavefunction in the momentum space
\begin{equation}
FT[\psi(q)](p) = \psi^{\rm FT}(p) = \psi(p)  = \frac{1}{2\sqrt{\pi}}  \int dq \psi(q) e^{-i q p/2}.
\end{equation}
An important class of wavefunctions are the squeezed-displaced wavefunctions, which are given by
\begin{equation}\label{eqWavefunctionG}
G_{x_0,s}(x) := \frac{1}{\sqrt{s}\pi^{1/4}} e^{-\frac{(x-x_0)^2}{2 s^2}},
\end{equation}
with squeezing $s>0$ and displacement $x_0$. We can simply check with $\int dx e^{-a (x-x_0)^2} = \sqrt{\pi/a}$ that these wavefunctions are normalized:
$\int dx \,\, G_{x_0,s}(x) G_{x_0,s}^\ast(x) =  \frac{1}{s \sqrt{\pi}} \int dx \,\, e^{- \frac{(x-x_0)^2}{s^2}} = \frac{1}{s \sqrt{\pi}} s \sqrt{ \pi}=1$.
The Fourier transform of the Gaussian is $FT[e^{-a x^2}](p) = \sqrt{\frac{\pi}{a}} e^{-\pi^2 p^2/ a}$.
The wavefunction in momentum space is found via
\begin{equation}
FT[G_{0,s}(x)](p) =  \frac{1}{\sqrt{s}\pi^{1/4}} s \sqrt{2 \pi} e^{-2 \pi^2 s^2 p^2} = \sqrt{2 s}\pi^{1/4} e^{-2 \pi^2 s^2 p^2} = G_{0,\frac{1}{2\pi s}}(p).
\end{equation}
Checking the normalization obtains $2 s \pi^{1/2} \int dp \,\, e^{-4 \pi^2 s^2 p^2} = 2 s \pi^{1/2} \sqrt{\pi/ (4 \pi^2 s^2)} = 1$.
Note that the vacuum is given by $s=1/\sqrt{2}$
\begin{equation}
\ket{\rm vac} = \int dx \,\, G_{0,\frac{1}{\sqrt{2}}}(x) \ket x_q. 
\end{equation}
These squeezed-displaced states can be generated by unitary operations applied to the vacuum. 
Define the displacement operator  \cite{Braunstein2005,Weedbrook2012} 
\begin{equation}
D(\alpha) := e^{\alpha \hat a^\dagger - \alpha^\ast \hat a},
\end{equation}
with parameter $\alpha \in \mathbbm C$.
Define the squeezing operator
\begin{equation} \label{eqSqueezeOp}
S(r) = e^{\frac{r}{2}( \hat a^2 - \hat a^{\dagger 2})},
\end{equation}
where $r \in \mathbbm R$ is a parameter determining the amount of squeezing. We take $r$ to be real in this work. If $r >0$ the $\hat q$ quadrature is squeezed and the $\hat p$ quadrature is anti-squeezed, and vice versa for $r < 0$.
Applying the squeezing operator to the vacuum leads to the above-defined wavefunctions of a squeezed state
\begin{equation}
S(r) D(x_0) \ket{\rm vac} = \left ( \frac{2}{\pi}\right )^{1/4} e^{r/2} \int dx \,\, e^{-e^{2r} (x-x_0)} \ket x = \int dx \,\, G_{x_0,s}(x) \ket x.
\end{equation}
We can relate 
\begin{eqnarray}\label{eqSqueezingFactorsRelation}
s=\frac{1}{\sqrt 2} e^{-r},\quad r= -\log (s \sqrt 2).
\end{eqnarray}
As mentioned before, $s=1/\sqrt{2}$, i.e. $r=0$, corresponds to the vacuum state. If $s<1/\sqrt{2}$ then  $r > 0$, which describes a position squeezed state. If $s>1/\sqrt{2}$ then  $r < 0$, which describes a position anti-squeezed (momentum squeezed) state. In the infinite squeezing limit we have the relation for the probability density
\begin{equation}
\lim_{s\to 0} \vert G_{x_0,s}(x) \vert^2 = \delta(x-x_0).
\end{equation}
In the infinite squeezing limit, also 
the wavefunction becomes proportional to the Dirac delta function, which defines the infinitely squeezed states
\begin{equation}
\ket x_q = \lim_{r\to \infty} D\left (x \right)S(r) \ket{\rm vac}.
\end{equation}
Finally, regarding norms, let $\left \vert \ket v \right \vert = \sqrt{\braket{v}{v}}$ be the norm of the quantum state $\ket v$. 
Given an operator $\mathcal U$, the operator norm is defined as
\begin{equation}
\Vert \mathcal U \Vert := \sup \left \{ \frac{\left \vert \mathcal U \ket v \right \vert }{ \vert \ket v \vert } : \ket v \neq 0\right \},
\end{equation}
which is the norm used in this work. 

\section{CV gate decompositions  and implementation}
\label{appendixDecomposition}

Decomposing arbitrary unitaries into elementary gates is a fundamental problem in quantum computing, both qubit and continuous-variable based. 
For continuous variables \cite{Sefi2011}, if the original unitary is Gaussian, that is at most second-order in the quadrature operators,
we can decompose it into multi-mode passive optics (beam splitters), single mode squeezing and displacements (Bloch-Messiah decomposition) \cite{Braunstein2005squeezing}. A single-mode Gaussian transformation can be decomposed into the Gaussian set 
$\{ e^{i\frac{\pi}{2} (\hat q^2 +\hat p^2)},e^{i t_1 \hat q},e^{i t_2 \hat p}\}$ with at most four steps \cite{Bartlett2002_2,Ukai2010}.
A single-mode arbitrary transformation can be decomposed into a universal CV set of gates for example given by 
$\{ e^{i\frac{\pi}{2} (\hat q^2 +\hat p^2)},e^{i t_1 \hat q},e^{i t_2 \hat q^2},e^{i t_3 \hat q^3}\}$. To achieve decompositions one can use unitary conjugation, approximation, linear combination simulation via the Lie product formula, and commutator simulation, to name a few \cite{Childs2002}. There are many relations regarding commutators of polynomials of the quadrature operators, we refer to \cite{Sefi2011}. 

The universal gate set includes the cubic phase gate $e^{i t_3 \hat q^3}$. An implementation of this gate was first discussed in \cite{Gottesman2001}. Applying this gate to an arbitrary state involves two ingredients \cite{Gottesman2001,Bartlett2002}. The first ingredient is the preparation of the cubic phase state $\ket \gamma = \int dx e^{i \gamma  x^3} \ket x$. Such a state can be prepared by preparing a two-mode squeezed state, a momentum kick in one of the modes, and a photon counting measurement in one of the modes. Conditioned on the outcome $n$, one then obtains a cubic phase state $\ket {\gamma'}$ with $\gamma' =\Ord{n^{-1/2}}$. With an additional squeezing operator one can turn $\gamma' \to \gamma$. The cubic phase states can be prepared offline and then used in the quantum algorithm as desired. This leads to the second ingredient, a gate teleportation of the cubic phase gate from the $\ket \gamma$ to an arbitrary state. The teleportation is achieved via first applying the gate $e^{i \hat q_1 \otimes \hat p_2}$ between the $\ket \gamma$ and the arbitrary state. Subsequently, a quadrature measurement in performed on the $\ket \gamma$ state and the other mode is adaptively shifted according to the outcome. This teleports the cubic phase gate over to the arbitrary state. 

In this work, we would like to integrate over the function $f(x)$ using the auxiliary polynomial $h(x)$ with degree $d_h$. For higher degrees, a decomposition of the polynomial into above universal set is prohibitive as the procedure is recursive. Going from a cubic term to a quartic term requires a number of operations, from quartic to quintic another number of operations, and so forth. For example \cite{Marek2018}, a polynomial of $10$-th order requires $4$ $6$-th order operations, each of which requires $16$ $4$-th order operations or $64$ $3$-rd order operations. In general, the number of cubic operations scales as $\Ord{2^{d_h}}$ with the degree of the polynomial. Ref.~\cite{Marek2018} presents 
a way of avoiding the recursive construction via directly implementing higher order gate via the merging of the adaptive operations. This results in a potential number of operations $\Ord{d_h}$, an exponential improvement in the degree of the polynomial. Alternatively, Ref.~\cite{Sabapathy2018} details a method of directly implementing higher order gates by using the ON resource state, a superposition of the vacuum and a fixed number of photons.

\section{Controlled rotation}
\label{appendixH}
In this appendix, we further elucidate the controlled rotation. The gate $\mathcal H$ for the controlled rotation is given in Eqs.~(\ref{eqGateHid}) and (\ref{eqGateHimpl}). Using the resolution of identity $\int dx \ket x \bra x = \mathbbm I$, we have
\begin{eqnarray}
 \mathcal H^{\rm impl} \ket x_{q_1} \ket{0}_{q_2} \ket 0_{q_3} &=& \ket x_{q_1}  \int dp'  \ket{p'}_{p_2} \ket { h(x) p'}_{q_3}\\
 &=& \ket x_{q_1}  \int dp' \int dx_2 \int dx_3 \ket{x_2}_{q_2} \braket{x_2}{p'}_{q_2,p_2} \ket{x_3}_{q_3} \braket{x_3} { h(x) p'}_{q_3}\\
  &=&  \frac{1}{2\sqrt{\pi}}  \ket x_{q_1}  \int dp' \int dx_2 \int dx_3 \ket{x_2}_{q_2} e^{+i p' x_2} \ket{x_3}_{q_3} \delta( h(x) p'-x_3) \\  
  &=&  \frac{1}{2\sqrt{\pi} \vert h(x)\vert }  \ket x_{q_1} \int dx_2 \int dx_3  e^{+i\frac{x_3 x_2}{h(x)}} \ket{x_2}_{q_2}\ket{x_3}_{q_3}.
\end{eqnarray}
We have used that $ \braket{x_2}{p'}_{q_2,p_2} =  \frac{e^{+i p' x_2} }{2\sqrt{\pi}} $ and 
$\braket{x_3} { h(x) p'}_{q_3}= \delta( h(x) p'-x_3)$. 
Here, we have also used the simple identity for the Dirac delta distribution
\begin{equation}
\delta(a x) = \frac{\delta(x)}{\vert a \vert}.
\end{equation}
Indeed,
\begin{eqnarray}
\delta( x_3- h(x) p') &=& \delta \left(  h(x) \left (\frac{x_3}{ h(x)} - p' \right)\right )\\
& =& \frac{1}{ \vert  h(x)\vert }\delta \left (\frac{x_3}{ h(x)} - p' \right).
\end{eqnarray} 
 This is a continuous variable analogue of the qubit controlled rotation. Let $0 < \sqrt{f(x)} < 1$ and 
$\ket{x} $ represent a qubit state with a binary encoding of $x$. Then the qubit controlled rotation is 
\begin{equation}
\ket{x} \ket 0 \to \ket{x} \left ( \sqrt{1-f(x)} \ket 0 + \sqrt{f(x)} \ket 1 \right ).
\end{equation}
We see that the CV controlled rotation involves an integral over all ancilla states while the qubit rotation involves a summation over the ancilla states, both with each term having a different amplitude. A projective measurement then arrives at the desired outcome. In the qubit case, 
measuring the ancilla in $\ket 1$ obtains a state $\propto\sqrt{ f(x)}\ket{x}$.
Applying $\bra{0}_{q_2}\otimes \bra{x_{\rm off}}_{q_3}$ to the CV state results in
\begin{eqnarray}
&&\left (\bra{0}_{q_2}\otimes \bra{x_{\rm off}}_{q_3}\right ) \ket x_{q_1}  \int dp'  \ket{p'}_{p_2} \ket { h(x) p'}_{q_3}\\
&&\quad = \ket x_{q_1}  \int dp'  \braket{0}{p'}_{q_2,p_2} \braket {x_{\rm off}}{ h(x) p'}_{q_3}\\
&&\quad = \ket x_{q_1}  \int dp'  \braket{0}{p'}_{q_2,p_2} \delta( x_{\rm off} -  h(x) p')\\
&&\quad = \ket x_{q_1}  \frac{1}{\vert  h(x) \vert}\braket{0}{\frac{x_{\rm off}}{h(x)}}_{q_2,p_2} 
= \frac{1}{2\sqrt{\pi} \vert  h(x) \vert} \ket x_{q_1},
\end{eqnarray}
where $\braket{0}{\frac{x_{\rm off}}{ h(x)}}_{q_2,p_2} = e^{i 0\frac{x_{\rm off}}{ h(x)}}/(2\sqrt{\pi}) =  1/(2\sqrt{\pi}) $.

\section{Error due to polynomial approximation}
\label{appendixErrorPolynomial}
Let $p(x) >0$ be a probability distribution. Let $ \bar{\mathcal X_h}$ be the complement of $\mathcal X_h$.
The polynomial approximation error on the complement is $\int_{  \bar{\mathcal X_h}} dx p(x) \left \vert \frac{1}{\vert  h(x) \vert^2}  - f(x) \right \vert $ and we assume that this error is $\leq \eta$. 
The assumption amounts to the function $p(x)$ being negligible in the complement $ \bar{\mathcal X_h}$, for example it has exponentially suppressed tails that are much smaller than the polynomial approximation error $ \left \vert \frac{1}{\vert  h(x) \vert^2}  - f(x) \right \vert$. 
For the error in Eq.~(\ref{eqchiPchi}), it holds that
\begin{eqnarray}
&& \left \vert \int dx \frac{p(x)}{\vert  h(x) \vert^2}  - \int dx f(x)p(x) \right \vert \\
&&\quad \leq  
\int dx p(x) \left \vert \frac{1}{\vert  h(x) \vert^2}  - f(x) \right \vert 
\\
&& \quad\leq  
\epsilon_h \int_{\mathcal X_h} dx p(x)+\int_{ \mathcal X_h^\complement} dx p(x) \left \vert \frac{1}{\vert  h(x) \vert^2}  - f(x) \right \vert 
\\
&& \quad \leq  
\epsilon_h + \eta.
\end{eqnarray}
For evaluating the desired integral in the first stage, this error compounds with the error $\epsilon_{sq}$ from the finite squeezing
discussed in Appendix \ref{App:FiniteSqueezing}.

\section{Finite squeezing}\label{App:FiniteSqueezing}

\subsection{The squeezed-state projector}
\label{appendixProj}

Our main discussion on Monte Carlo integration via the amplitude estimation algorithm uses a projector into Gaussian squeezed states. The projector into the Gaussian squeezed states is defined as
\begin{equation}
P_{x_0,\Delta x} := \ket{G_{x_0,\Delta x} }\bra{G_{x_0,\Delta x} }.
\end{equation}
In this appendix, we analyze the difference of the Gaussian squeezed projector to the 
operator $\ket{x_0} \bra{x_0}$, which in a sense is the (unnormalized) limit as $\Delta x\to 0$. 
Note that with another squeezed state $\ket{G_{y,s} }$, we have
\begin{eqnarray}
\braket{G_{y,s} }{G_{x_0,\Delta x} } &=& \int dx G_{y,s}(x) G_{x_0,\Delta x}(x) dx \\
&=&\frac{\sqrt{2 s \Delta x} }{\sqrt{(\Delta x^2+s^2)}}e^{-\frac{(x_0-y)^2}{2(\Delta x^2+s^2)}}.
\end{eqnarray}
and thus
\begin{eqnarray}
\bra{G_{y,s} } P_{x_0,\Delta x} \ket{G_{y,s} } &=& \braket{G_{y,s} }{G_{x_0,\Delta x} }\braket{G_{x_0,\Delta x} }{G_{y,s} }\\
&=&\frac{2 s \Delta x }{\Delta x^2+s^2}e^{-\frac{(x_0-y)^2}{\Delta x^2+s^2}}\\
&=& \frac{2 \Delta x }{s}e^{-\frac{(x_0-y)^2}{s^2}} + \Ord{\Delta x^2}.
\end{eqnarray}
In contrast, using the position eigenstates
\begin{equation}
\braket{G_{y,s} } {x_0} \braket{x_0}{G_{y,s} } = \frac{e^{-\frac{(x_0-y)^2}{s^2}}}{\sqrt{\pi} s }.
\end{equation}
Thus, as expected, the Gaussian projector does not have the same limit as the unnormalized operator $\ket {x_0} \bra{x_0}$. In the infinite squeezing limit, we obtain the conversion factor $\frac{1}{2\sqrt{\pi} \Delta x}$, which reappears in the more involved calculations below. 

\subsection{Realistic projector and finitely squeezed initial states} \label{appendixFiniteSqueezing}

As discussed before, the ancilla computational states we used to derive Eqns.~(\ref{eqHPostSelectSuperposition}) and (\ref{eqHPostSelectProb}) are infinitely squeezed. 
We have assumed that we can prepare
 \begin{equation}
 \int dx \sqrt{p(x)} \ket x_{q_1}  \ket{0}_{q_2} \ket 0_{q_3},
\end{equation}
where the last two modes are infinitely squeezed.
In this appendix, we consider both the realistic projector and finitely squeezed initial states and derive Eq.~(\ref{eqchiPchi}). Take $r_{\max}$ the maximum achievable squeezing factor with corresponding squeezing $s_{\rm min}$. To this end, 
we apply the projector $\mathcal P :=\mathbbm I_{1}\otimes P_{0,s_{\rm min}} \otimes P_{x_{\rm off},s_{\rm min}}$ to the state $\ket {\chi^{\rm impl}} = \mathcal H^{\rm impl} \left (\mathcal G\otimes S(r_{\max}) \otimes S(r_{\max}) \right )\ket {\psi_{\rm in}} $. We then apply $\bra {\chi^{\rm impl}}$ to the result to obtain Eq.~(\ref{eqchiPchi}).
We abbreviate $s_{\rm min}$ with $s$ in the following intermediate steps. 
Applying $\mathcal G\otimes S(r_{\max}) \otimes S(r_{\max})$ to the vacuum initial states obtains
 \begin{equation}
\int \int \int dx_1 dx_2 dx_3 \sqrt{p(x_1)} G_{0,s} (x_2) G_{0,s}(x_3)\ket {x_1}_{q_1}  \ket{x_2}_{q_2} \ket {x_3}_{q_3},
\end{equation}
with $G_{0,s}(x) = \frac{1}{\sqrt{s}\pi^{1/4}} e^{-x^2/2 s^2}$ as in Eq.~(\ref{eqWavefunctionG}).
Applying the gate $ \mathcal H^{\rm impl}$ obtains the following state
 \begin{equation}
\int \int \int dx_1 dx_2 dx_3 \int dp_2 \sqrt{p(x_1)} G_{0,s} (x_2) G_{0,s}(x_3) \ket {x_1}_{q_1}  e^{-i p_2 x_2}\ket{p_2}_{p_2} \ket {x_3+  h(x_1)p_2}_{q_3} =: \ket {\chi^{\rm impl}}.
\end{equation}
Applying the projector gives 
\begin{eqnarray}
\mathcal P\ket {\chi^{\rm impl}} &=& \mathbbm I_{1}\otimes P_{0,s} \otimes P_{x_{\rm off},s}
\int \int \int dx_1 dx_2 dx_3 \int dp_2 \sqrt{p(x_1)} G_{0,s} (x_2) G_{0,s}(x_3) \ket {x_1}_{q_1}  e^{-i p_2 x_2}\ket{p_2}_{p_2} \ket {x_3+  h(x_1)p_2}_{q_3} \nonumber \\
&=& \int dx_2'' G_{0,s} (x_2'') \int dx_3'' G_{x_{\rm off},s} (x_3'')
\int \int \int dx_1 dx_2 dx_3 \int dp_2 \times \\
&& \times  \sqrt{p(x_1)} G_{0,s} (x_2) G_{0,s}(x_3) e^{-i p_2 x_2}e^{+i p_2 x_2''} \delta (x_3+  h(x_1)p_2 -x_3'') \ket {x_1}_{q_1}  \ket{G_{0,s} } \ket{G_{x_{\rm off},s} }\nonumber \\
&=& \int dx_2'' G_{0,s} (x_2'') \int dx_3'' G_{x_{\rm off},s} (x_3'')
\int \int \int dx_1 dx_2 dx_3 \times \\
&& \times  \sqrt{p(x_1)} G_{0,s} (x_2) G_{0,s}(x_3) e^{-i \frac{x_3-x_3''}{h(x_1)} (x_2 - x_2'')}\frac{1}{\vert h(x_1)\vert} \ket {x_1}_{q_1}  \ket{G_{0,s} } \ket{G_{x_{\rm off},s} }\nonumber \\
&=& \int dx_1\frac{ \sqrt{p(x_1)} }{\vert h(x_1)\vert} B(x_1,s,x_{\rm off}) \ket {x_1}_{q_1}  \ket{G_{0,s} } \ket{G_{x_{\rm off},s} }.
\end{eqnarray}
Here, we have defined
\begin{eqnarray}
B(x_1,s,x_{\rm off}) &:=& \int  \int 
\int \int  dx_2 dx_3 dx_2'' dx_3'' G_{0,s} (x_2) G_{0,s}(x_3) G_{0,s} (x_2'') G_{x_{\rm off},s} (x_3'')e^{-i \frac{(x_3-x_3'')(x_2 - x_2'')}{\vert  h(x_1)\vert^2} } \\
&=& \frac{4 e^{-\frac{2 s^2 x_{\rm off}^2}{\vert  h(x_1)\vert^2+4 s^4}}}{\frac{4}{\vert  h(x_1)\vert^2}+\frac{1}{s^4}}.
\end{eqnarray}
Furthermore, we have
\begin{eqnarray}
\bra {\chi^{\rm impl}} \mathcal P \ket {\chi^{\rm impl}} &=& \int dx_1 \frac{p(x_1)}{\vert  h(x_1)\vert^2}  B^2(x_1,s,x_{\rm off}).
\end{eqnarray}
In the limit of infinite squeezing $s \to 0$, we can expand
\begin{eqnarray}
B^2(x_1,s,x_{\rm off})  &=&
  4 s^4 - \frac{8 x_{\rm off}^2 s^6}{\vert  h(x_1)\vert^2}+\Ord{s^8}.
\end{eqnarray}
Define the infinitely-squeezed limit
\begin{equation}
B^2_{\infty}  :=  4 s^4.
\end{equation}
We provide an error estimate. As we have the function $B$ in the integral we estimate how far the integral is from the desired integral. 
We evaluate the integrated difference to the infinite-squeezing limit $\epsilon_{sq}$ to be
\begin{eqnarray}
\epsilon_{sq} &\leq& \int dx_1 \frac{p(x_1)}{\vert  h(x_1)\vert^2} \left \vert B^2(x_1,s,x_{\rm off}) - B^2_{\infty} \right \vert \\
&\leq& \int dx_1 \frac{p(x_1)}{\vert  h(x_1)\vert^2}  \left \vert \Ord{\frac{ x_{\rm off}^2 s^6}{\vert  h(x_1)\vert^2} } \right \vert \\
&=&\Ord{s^6 \mathbbm V[1/\vert  h(x_1)\vert^2 ] x_{\rm off}^2 }.
\end{eqnarray}
The error is proportional to the variance of the random variable and $s^6$.

\section{Reflection operators}
\label{SecReflec}

\subsection{The PBL reflection}
\label{appendixPBL}

In this section, we discuss the reflection used in Ref.~\cite{Pati2000} for Grover search with continuous variables. We name this reflection the PBL reflection after the authors Pati, Braunstein, and Lloyd. 
The task of this gate is to assign a negative phase to certain position eigenstates. Let $\ket {x_0}$ be such a basis state. Then the operation is 
\begin{equation}
\ket {x_0} \to -\ket {x_0},
\end{equation}
and the identity operation otherwise. Let $\Delta x >0$, the task is to implement the operation
\begin{equation} \label{eqPBLIdeal}
\mathcal C_{x_0,\Delta x}^{\rm PBL} = \mathbbm I - 2 P_{x_0,\Delta x}^{\rm PBL}.
\end{equation}
with 
\begin{equation}
P_{x_0,\Delta x}^{\rm PBL}  = \int_{x_0 -\Delta x/2}^{x_0 +\Delta x/2} dx \ket x \bra x.
\end{equation}
This projector is different from the one used in the main part of the paper Eq.~(\ref{eqProj}).

To implement this operation, first, let a function be $c_{x_0, \Delta x}: \mathbbm R \to \{0,1\}$ with $c_{x_0, \Delta x}(x) = 1$ if $x=[x_0-\frac{\Delta x}{2}, x_0+\frac{\Delta x}{2}]$, and $c_{x_0, \Delta x}(x) = 0$ otherwise. 
Replacing operator $\hat q$ for $x$,
the operator $c_{0, \Delta x}(\hat q)$ picks out the zero-position component of a state with a spread $\Delta x$. 
Assume we can implement the two-mode gate
\begin{equation}\label{eqCImpl}
\mathcal C_{0,\Delta x}^{\rm PBL, impl} :=e^{-i c_{0,\Delta x}(\hat q_1)\otimes \hat p_z}.
\end{equation}
The mode $z$ is an ancilla mode for this operation.  
Applying $\mathcal C_{0,\Delta x}^{\rm PBL, impl} $ to a computational state leads to 
\begin{equation}
\ket{x}_{q_1} \ket{a}_{q_z} \to \ket{x}_{q_1} \ket{c_{0,\Delta x}(x)+a}_{q_z}.
\end{equation}
To perform the desired operation, apply this quantum gate with ancilla state $\ket{f}_{q_z} := \mathcal F \ket{\pi/2}_{q_z} = \int dy e^{i\pi y} \ket y$, where $\mathcal F$ is the Fourier transform. This achieves
\begin{eqnarray}
\ket{x}_{q_1} \ket{f}_{q_z}  &=& \ket{x}_{q_1} \int dy e^{i\pi y} \ket y \\
&\to&\ket{x}_{q_1}  \int dy e^{i\pi y} \ket {c_{0,\Delta x}(x)+ y} \\
&=&\ket{x}_{q_1} \int dy e^{i\pi (y-c_{0,\Delta x}(x))} \ket {y} \\
&=&e^{-i \pi c_{0,\Delta x}(x)} \ket{x}_{q_1} \ket{f}_{q_z} .
\end{eqnarray}
If $x=0$ with an uncertainty $\Delta x$, this operation obtains a phase $-1$, otherwise the phase is $+1$, as desired.

We continue with defining the gate error. The main source of error is from the implementation of the function $c_{0,\Delta x}(x)$, which for example has to be approximated via a polynomial. 
The error arising from using the exact gate $\mathcal C_{x_0,\Delta x}^{\rm impl} $, see Eq.~(\ref{eqCImpl}), is
by definition 
\begin{equation}
\Vert \mathcal C_{x_0,\Delta x}^{\rm PBL}   -  \bra{f}_{q_z} \mathcal C_{x_0,\Delta x}^{\rm PBL, impl} \ket{f}_{q_z}\Vert = 0,
\end{equation}
where the reduced operator in the first register is compared by keeping the ancilla register for the $\mathcal C_{x_0,\Delta x}^{\rm impl}$ gate in the $\ket{f}_{q_z}$ state. This zero error holds if we can implement the function $c_{x_0,\Delta x}$ exactly. 
As we are usually implementing the function $c_{x_0,\Delta x}$ inexactly via a gate decomposition, let 
\begin{equation}
\mathcal C_{x_0,\Delta x}^{\rm PBL, dec} 
\end{equation}
be the CV gate decomposition of $\mathcal C_{x_0,\Delta x}^{\rm PBL, impl} $. Define the error of this decomposition as 
\begin{equation}
\epsilon_C^{\rm dec} := \left \Vert \bra{f}_{q_z} \left( \mathcal C_{x_0,\Delta x}^{\rm dec} - \mathcal C_{x_0,\Delta x}^{\rm impl}  \right) \ket{f}_{q_z} \right \Vert.
\end{equation}
We leave a quantification of this error for future work. 
In summary, the total error of the PBL gate is given by
\begin{equation}
\epsilon_{\rm PBL} := \left \Vert  \mathcal C_{x_0,\Delta x}^{\rm PBL} -\bra{f}_{q_z} \mathcal C_{x_0,\Delta x}^{\rm dec} \ket{f}_{q_z} \right \Vert = \Ord{\epsilon_C^{\rm dec} }.
\end{equation}

\subsection{The vacuum reflection operator}
\label{appendixVacuum}

The qubit amplitude estimation uses a reflection around the computational initial state of all qubits being in the $\ket 0$ state. In the CV case the initial state is the zero-photon vacuum state. For the CV version of amplitude estimation, we require a reflection around such a state. 
We first discuss the single-mode vacuum reflection operator. It is given by 
\begin{equation}
\mathcal Z_1^{\rm id} = \mathbbm I - 2 \ket{\rm vac}\bra{\rm vac}.
\end{equation}
This reflection operator puts a phase $-1$ to the optical ground state and leaves all other photon states unchanged. 
The operator can be expressed in the Fock basis $\ket{k}_{n}$ as
\begin{equation}
\mathcal  Z_1^{\rm id} = -  \ket{0}_{n} \bra{0}_{n} + \sum_{k=1}^\infty  \ket{k}_{n} \bra{k}_{n}.
\end{equation}
Note that $\mathcal Z_1^{\rm id}$ can also be written as
\begin{equation}
\mathcal  Z_1^{\rm id} = e^{i \pi \ket{0}_{n} \bra{0}_{n} }.
\end{equation}
One way to implement the Hamiltonian $\ket{0}_{n} \bra{0}_{n}$ may be to express it as a polynomial of 
the quadratures $\hat q$ and $\hat p$. Here, we show an implementation via the PBL reflection discussed in Appendix \ref{appendixPBL}. 
While this technique formally allows to implement the vacuum reflection, there may exist simpler and more realistic methods. After all $e^{i \pi \ket{0}_{n} \bra{0}_{n} }$ is a simple operation that attaches a phase $-1$ to the ground state and is identity on all other states, which may be accomplished straightforwardly in an actual physical system. 
The equivalent operation for qubits has a straightforward implementation \cite{Montanaro2015,Xu2018}. 

\subsection{The vacuum reflection via the PBL reflection}

Here, we discuss an implementation of the vacuum reflection that employs the PBL gate of Appendix \ref{appendixPBL}. 
First, we can define
\begin{equation}
 \mathcal Z_1^{\rm id'} := \mathbbm I - \frac{2}{\Delta x} \int_{-\Delta x/2}^{\Delta x/2} dx\ D(x) \ket{\rm vac}\bra{\rm vac} D^\dagger(x) =:  \mathbbm I - 2 \frac{ \mathcal B}{\Delta x},
\end{equation}
which is close to $\mathcal Z_1^{\rm id}$. 
This can be seen with $k,k'\geq1$ from 
\begin{eqnarray}
\bra {\rm vac} \mathcal Z_1^{\rm id'} \ket {\rm vac} &=& 1 - \frac{2}{\Delta x}  \int_{-\Delta x/2}^{\Delta x/2} dx\ e^{-x^2} = -1 + \Ord{\Delta x^2} \\
\bra {k}_n \mathcal Z_1^{\rm id'} \ket {k}_n &=& 1 - \frac{2}{\Delta x}  \int_{-\Delta x/2}^{\Delta x/2} dx\ e^{-x^2} \frac{x^{2k}}{k!} = 1 + \Ord{\Delta x^2} \\
\bra {k'}_n \mathcal Z_1^{\rm id'} \ket {k}_n &=& - \frac{2}{\Delta x}  \int_{-\Delta x/2}^{\Delta x/2} dx\ e^{-x^2} \frac{x^{k+k'}}{\sqrt{k!k'!}} =  \Ord{\Delta x^2}.
\end{eqnarray}
Thus, we have 
\begin{equation}
\left \Vert  \mathcal Z_1^{\rm id'}  -  \mathcal Z_1^{\rm id} \right \Vert = \Ord{\Delta x^2}.
\end{equation}
In addition, we can
expand the projector used for the PBL gate
$\mathcal C_{0,\Delta x} = \mathbbm I - 2 P_{0,\Delta x}^{\rm PBL}$, i.e.,
\begin{eqnarray}
P_{0,\Delta x}^{\rm PBL} &=& \int_{-\Delta x/2}^{+\Delta x/2} dx \ket x \bra x\\
&=& \int_{-\Delta x/2}^{+\Delta x/2} dx S(\infty) D(x) \ket{\rm vac} \bra {\rm vac} D^\dagger(x) S(\infty)^\dagger = S(\infty) \mathcal B S(\infty)^\dagger.
\end{eqnarray}
Consider that the infinite squeezing operators are formally defined, which accounts for the $\frac{1}{\Delta x}$ factor. Thus, we note the following relationship
\begin{equation}
\mathcal Z_1^{\rm id} \approx S(\infty)^\dagger  \mathcal C_{0,\Delta x}^{\rm PBL} S(\infty ), 
\end{equation}
which holds up to error $\Ord{\Delta x^2}$.
In words, we are squeezing strongly such that the vacuum is squeezed into a state with 
position spread below $\Delta x$ and all other states are squeezed accordingly. Then, we apply the PBL  gate $\mathcal C_{0,\Delta x}^{\rm PBL}$, which attaches a phase $-1$ to all position states 
in the interval  $[-\frac{\Delta x}{2},\frac{\Delta x}{2}]$. 
Then, we unsqueeze position, i.e., squeeze in the momentum component, via $S(\infty)^\dagger$ to undo the initial squeezing and obtain a state where the vacuum component receives a phase $-1$. As long as the squeezing factor is large enough, this operation obtains the desired reflection of the optical ground state, at least formally. Together with the error $\epsilon_S$ from finite squeezing $S(\infty)\to S(r_{\max})$, see Eq.~(\ref{eqSError}), we obtain
the error $\Ord{\Delta x^2 + \epsilon_S}$.

\subsection{The reflection operator $\mathcal Z$}
\label{appendixZ}

The operator $\mathcal Z$ is a reflection around the computational initial state given 
by
\begin{equation}
\mathcal Z \equiv \mathcal Z^{\rm id}_3 =  
 \mathbbm I ^{\otimes 3} - 2 \ket{\rm vac}_{1} \bra{\rm vac}_{1}\otimes \ket{\rm vac}_{2} \bra{\rm vac}_{2} \otimes  \ket{\rm vac}_{3} \bra{\rm vac}_{3} = e^{-i\pi \ket{\rm vac}_{1} \bra{\rm vac}_{1}\otimes \ket{\rm vac}_{2} \bra{\rm vac}_{2} \otimes  \ket{\rm vac}_{3} \bra{\rm vac}_{3}}.
\end{equation}
It is implemented analogous to the single-mode vacuum reflection. 
The `ideal' reflection is equivalently defined via a spread $\Delta x'$ for which the computational states are equivalent. 
Note the relationship 
\begin{equation}
\mathcal Z_3^{\rm id} \approx S(r_{\Delta x'} )^\dagger \otimes S(r_{\Delta x'} )^\dagger \otimes S(r_{\Delta x'} )^\dagger \left( \mathcal C_{0,\Delta x}^{\rm PBL} \otimes \mathcal C_{0,\Delta x}^{\rm PBL} \otimes \mathcal C_{0,\Delta x}^{\rm PBL} \right) S(r_{\Delta x'} ) \otimes S(r_{\Delta x'} ) \otimes S(r_{\Delta x'} ),
\end{equation}
which holds again up to order $\Ord{\Delta x^2}$. 
Here, we squeeze all three modes accordingly and use the PBL gate for each of the three modes. 
Replacing $ S(r_{\Delta x'} ) \to S(r_{\max})$  incurs an additional error 
$\epsilon_S$.
This shows the reflection around the computational initial state used for the continuous-variable amplitude amplification algorithm.

\subsection{The reflection operator $\mathcal V$}
\label{appendixV}

For the amplitude amplification algorithm, we also require a reflection $\mathcal V$ defined by the measurement projector. 
The `ideal' reflection is equivalently defined via a spread $\Delta x'$ for which the computational states are equivalent. It is given by
\begin{eqnarray}
\mathcal V^{\rm id} &=& \mathbbm I^{\otimes 3} -2 \mathbbm I \otimes P_{0,\Delta x'}  \otimes P_{x_{\rm off},\Delta x'}\\
&\equiv&\mathbbm I^{\otimes3} -2 \mathbbm I \otimes \ket{G_{0,\Delta x'} }_2 \bra{G_{0,\Delta x'} }_2 \otimes \ket{G_{x_{\rm off},\Delta x'} }_3 \bra{G_{x_{\rm off},\Delta x'} }_3\\
&=&\left(\mathbbm I \otimes S(r_{\Delta x'}) \otimes D(x_{\rm off}) S(r_{\Delta x'}) \right) \left( \mathbbm I^{\otimes3} -2 \mathbbm I \otimes \ket{\rm vac }_2\bra{\rm vac}_2\otimes \ket{\rm vac }_3\bra{\rm vac }_3 \right ) \times \nonumber \\
&& \times \left ( \mathbbm I \otimes S^\dagger(r_{\Delta x'}) \otimes S^\dagger(r_{\Delta x'})  D^\dagger(x_{\rm off})  \right).
\end{eqnarray}
Thus, the gate can be implemented by squeezing, displacement, and a two-mode vacuum reflection. Such a reflection is implemented analogous to Appendix \ref{appendixZ}.

\section{Details on the Grover operator $\mathcal Q$ and its controlled version}
\label{appendixQ}

The operator $\mathcal Q^{\rm id}$, defined in Eq.~(\ref{eqQ}), preserves a two-dimensional subspace. 
Dropping some $^{\rm id}$ superscripts, we have
\begin{eqnarray}
\mathcal Q^{\rm id} \mathcal V \ket{\chi} &=& -(\mathbbm I -2 \ket \chi \bra \chi) \mathcal V \ket{\chi} =-\mathcal V \ket \chi + 2\cos  (\theta/2) \ket {\chi}, \nonumber
\end{eqnarray}
and
\begin{eqnarray}
\mathcal Q^{\rm id} \ket{\chi} &=&  (\mathbbm I -2 \ket \chi \bra \chi) (\ket \chi - 2 \cos  (\theta/2) \mathcal V \ket \chi)\nonumber \\
&=& \ket \chi - 2 \cos  (\theta/2) \mathcal V \ket \chi -2 \ket \chi + 4  \cos^2  (\theta/2) \ket \chi \nonumber\\
&=& (4\cos^2  (\theta/2) - 1) \ket \chi - 2 \cos  (\theta/2) \mathcal V \ket \chi.\nonumber
\end{eqnarray}
We see that the result of applying $\mathcal Q^{\rm id}$ to $\ket \chi$ and $\mathcal V \ket \chi$ is preparing linear combinations of these two states.
In addition, we can express $\mathcal Q^{\rm id}$ with respect to the orthogonal states $\ket \chi$ and $\ket {\chi^\perp}$ as
\begin{eqnarray}\label{eqQtwobytwo}
\mathcal Q^{\rm id} \ket{\chi} &=&  (2\cos^2  (\theta/2) - 1) \ket \chi  - 2 \cos(\theta/2) \sin(\theta/2) e^{i\phi} \ket{\chi^\perp},\\
\mathcal Q^{\rm id} \ket {\chi^\perp} &=& 2 \cos(\theta/2) \sin(\theta/2) e^{-i\phi} \ket{\chi} +(2\cos(\theta/2)-1) \ket{\chi^\perp},\nonumber
\end{eqnarray}
where $\frac{2 \cos(\theta/2) - 2\cos^3(\theta/2)}{\sin(\theta/2)} = 2 \cos(\theta/2)\sin(\theta/2)$ was used. 
Diagonalizing the matrix associated with Eqs.~(\ref{eqQtwobytwo}) leads to the eigenvalues $e^{\pm \theta}$, with
associated eigenstates
\begin{eqnarray}
\ket{\psi_\pm} &=&\frac{1}{\sqrt{2}} \left( \ket \chi \pm i \ket {\chi^\perp} \right).
\end{eqnarray}

We now show the sequence of gates making up $\mathcal Q^{\rm id}$. For the probability distribution $p(x)$, we have the operator $\mathcal G$ which we leave as a variable in the following. Define here $\mathcal G' :=\mathcal G\otimes S(r_{\Delta x'}) \otimes S(r_{\Delta x'})$. 
We write out explicitly the operator $\mathcal Q^{\rm id} $ as
\begin{eqnarray}\label{Eq:BigQid}
\mathcal Q^{\rm id} &=&  \mathcal H^{\rm id}  \mathcal G' \mathcal Z^{\rm id}  \mathcal G'^\dagger (\mathcal H^{\rm id} )^\dagger \mathcal V^{\rm id} \mathcal H^{\rm id}  \mathcal G' \mathcal Z^{\rm id} \mathcal G'^\dagger (\mathcal H^{\rm id} )^\dagger \mathcal V^{\rm id}\\
&=&  e^{-i 1/\sqrt{f(\hat q_1)}\otimes \hat p_2 \otimes \hat p_3}  \left (\mathcal G\otimes S(r_{\Delta x'}) \otimes S(r_{\Delta x'}) \right ) e^{-i\pi \ket{\rm vac}_{1} \bra{\rm vac}_{1}\otimes \ket{\rm vac}_{2} \bra{\rm vac}_{2} \otimes  \ket{\rm vac}_{3} \bra{\rm vac}_{3}}  \\
&& \left (\mathcal G\otimes S(r_{\Delta x'}) \otimes S(r_{\Delta x'}) \right )^\dagger e^{i 1/\sqrt{f(\hat q_1)}\otimes \hat p_2 \otimes \hat p_3} \nonumber\\
&&\left(\mathbbm I \otimes S(r_{\Delta x'}) \otimes D(x_{\rm off}) S(r_{\Delta x'}) \right) e^{i \pi \mathbbm I\otimes \ket{\rm vac}_{2} \bra{\rm vac}_{2} \otimes  \ket{\rm vac}_{3} \bra{\rm vac}_{3} } \left ( \mathbbm I \otimes S^\dagger(r_{\Delta x'}) \otimes S^\dagger(r_{\Delta x'})  D^\dagger(x_{\rm off})  \right) \nonumber \\
 && e^{-i 1/\sqrt{f(\hat q_1)}\otimes \hat p_2 \otimes \hat p_3}  \left (\mathcal G\otimes S(r_{\Delta x'}) \otimes S(r_{\Delta x'}) \right ) e^{-i\pi \ket{\rm vac}_{1} \bra{\rm vac}_{1}\otimes \ket{\rm vac}_{2} \bra{\rm vac}_{2} \otimes  \ket{\rm vac}_{3} \bra{\rm vac}_{3}} \nonumber \\
 && \left (\mathcal G\otimes S(r_{\Delta x'}) \otimes S(r_{\Delta x'}) \right )^\dagger e^{i 1/\sqrt{f(\hat q_1)}\otimes \hat p_2 \otimes \hat p_3}\nonumber  \\
&& \left(\mathbbm I \otimes S(r_{\Delta x'}) \otimes D(x_{\rm off}) S(r_{\Delta x'}) \right) e^{i \pi \mathbbm I\otimes \ket{\rm vac}_{2} \bra{\rm vac}_{2} \otimes  \ket{\rm vac}_{3} \bra{\rm vac}_{3} } \left ( \mathbbm I \otimes S^\dagger(r_{\Delta x'}) \otimes S^\dagger(r_{\Delta x'})  D^\dagger(x_{\rm off})  \right).\nonumber
\end{eqnarray}
The controlled version is
\begin{eqnarray}\label{Eq:BigQcid}
\mathcal Q^{\rm id}_c 
&=&  e^{-i 1/\sqrt{f(\hat q_1)}\otimes \hat p_2 \otimes \hat p_3 \otimes \hat p_\phi} \left (\mathcal G_c\otimes S_c(r_{\Delta x'}) \otimes S_c(r_{\Delta x'}) \right ) e^{-i\pi \ket{\rm vac}_{1} \bra{\rm vac}_{1}\otimes \ket{\rm vac}_{2} \bra{\rm vac}_{2} \otimes  \ket{\rm vac}_{3} \bra{\rm vac}_{3} \otimes \hat p_\phi} \\
&& \left (\mathcal G_c\otimes S_c(r_{\Delta x'}) \otimes S_c(r_{\Delta x'}) \right )^\dagger e^{i 1/\sqrt{f(\hat q_1)}\otimes \hat p_2 \otimes \hat p_3\otimes \hat p_\phi} \\
&&\left(\mathbbm I \otimes S_c(r_{\Delta x'}) \otimes D_c(x_{\rm off}) S_c(r_{\Delta x'}) \right) e^{i \pi \mathbbm I\otimes \ket{\rm vac}_{2} \bra{\rm vac}_{2} \otimes  \ket{\rm vac}_{3} \bra{\rm vac}_{3} \otimes \hat p_\phi} \left ( \mathbbm I \otimes S^\dagger _c(r_{\Delta x'}) \otimes S^\dagger _c(r_{\Delta x'})  D^\dagger _c(x_{\rm off})  \right) \nonumber \\
 && e^{-i 1/\sqrt{f(\hat q_1)}\otimes \hat p_2 \otimes \hat p_3} \left (\mathcal G_c\otimes S_c(r_{\Delta x'}) \otimes S_c(r_{\Delta x'}) \right ) e^{-i\pi \ket{\rm vac}_{1} \bra{\rm vac}_{1}\otimes \ket{\rm vac}_{2} \bra{\rm vac}_{2} \otimes  \ket{\rm vac}_{3} \bra{\rm vac}_{3}\otimes \hat p_\phi} \nonumber \\
 &&\left (\mathcal G_c\otimes S_c(r_{\Delta x'}) \otimes S_c(r_{\Delta x'}) \right )^\dagger e^{i 1/\sqrt{f(\hat q_1)}\otimes \hat p_2 \otimes \hat p_3 \otimes \hat p_\phi}\nonumber  \\
&&\left(\mathbbm I \otimes S_c(r_{\Delta x'}) \otimes D_c(x_{\rm off}) S_c(r_{\Delta x'}) \right) e^{i \pi \mathbbm I\otimes \ket{\rm vac}_{2} \bra{\rm vac}_{2} \otimes  \ket{\rm vac}_{3} \bra{\rm vac}_{3}\otimes \hat p_\phi} \left ( \mathbbm I \otimes S^\dagger _c(r_{\Delta x'}) \otimes S^\dagger _{c}(r_{\Delta x'})  D^\dagger _{c}(x_{\rm off})  \right). \nonumber 
\end{eqnarray}
We use the controlled displacement and squeezing operators
\begin{eqnarray}
D_c(\alpha) &=& e^{ (\alpha\hat a^\dagger - \alpha^\ast \hat a) \otimes \hat p_\phi},\\
S_c(r) &=& e^{\frac{r}{2}( \hat a^2 - \hat a^{\dagger 2})\otimes \hat p_\phi}.
\end{eqnarray}

\section{Phase estimation}
\label{appendixPE}

The error of a single application of $\mathcal Q^c$ is given by $\epsilon_Q$, hence
the error of $ (\mathcal Q_c)^M$ is $\Ord{M\epsilon_Q}$.
This error appears as a spread $\Ord{M\epsilon_Q}$ of the $\theta$ values in the phase estimation mode. Thus, instead of 
state Eq.~(\ref{eqPhaseEstimationState}) we obtain the statistical mixture
\begin{eqnarray}
\frac{1}{s \sqrt \pi}\ket {\psi_+}\bra {\psi_+}    \int d \tilde\theta G^2_{M\theta,M \epsilon_Q} (\tilde \theta)
\int dx \int dx'  e^{-\frac{x^2+x'^2}{2s^2}} \ket {x+ \tilde \theta}_{q_\phi} \bra {x'+ \tilde \theta}_{q_\phi}.
\end{eqnarray}
Measuring the position $q$ obtains the probability distribution
\begin{eqnarray}
\tilde P_\theta(q) &=& \frac{1}{s \sqrt{\pi}} \int d \tilde \theta G^2_{M\theta,M \epsilon_Q} (\tilde \theta) \int dx \int dx' e^{-\frac{x^2 + x'^2}{2s^2}} \braket { x'+ \tilde\theta}{q} \braket{q} {x+ \tilde \theta}_{q_\phi} \\
&=& \frac{1}{s \sqrt{\pi}} \int d\theta' G^2_{M\theta,M \epsilon_Q} (\theta')e^{-\frac{(\tilde \theta-q)^2}{s^2}} \\
&=& \frac{e^{-\frac{ (M\theta-q)^2}{(M\epsilon_Q)^2+s^2} } }{{\sqrt \pi} \sqrt{(M\epsilon_Q)^2+s^2}}.
\end{eqnarray}
The success probability of $q$ being inside a range $M\epsilon$ around the expectation value $M\theta$, i.e., $\vert q - M\theta \vert \leq M\epsilon$, is given by
\begin{equation}
p_{\rm success} = \int_{M\theta -M\epsilon}^{M\theta + M\epsilon} dq \tilde P_\theta(q) = {\rm erf}\left( \frac{ M\epsilon}{\sqrt{(M\epsilon_Q)^2 +s^2}}\right).
\end{equation}

\end{document}